\documentclass[aps,pra,groupedaddress,nofootinbib,notitlepage,showpacs,floatfix]{revtex4-1}
\usepackage{amsfonts}

\usepackage{graphicx,graphics,epsfig,times,bm,bbm,amssymb,amsmath,amsfonts,mathrsfs}
\usepackage[normalem]{ulem}
\usepackage{wrapfig}
\usepackage{boxedminipage}
\usepackage{setspace}
\usepackage{subfigure}
\usepackage{dsfont}
\usepackage{braket}
\usepackage[pdftex]{color}
\usepackage[pdfstartview=FitH]{hyperref}

\newcommand{\bes} {\begin{subequations}}
\newcommand{\ees} {\end{subequations}}
\newcommand{\bea} {\begin{eqnarray}}
\newcommand{\eea} {\end{eqnarray}}
\newcommand{\beq} {\begin{equation}}
\newcommand{\eeq} {\end{equation}}

\newcommand{\mc}{\mathcal}

\def\b{\beta}

\def\>{\rangle}
\def\<{\langle}

\newcommand{\ketbra}[2]{|{#1}\>\<#2|}
\newcommand{\bracket}[2]{\<{#1}|{#2}\>}

\newcommand{\ident}{\mathbb{I}}
\newcommand{\ignore}[1]{}

\begin{document}

\author{Tameem Albash,$^{(1,2)}$ Sergio Boixo,$^{(2,3)}$ Daniel A. Lidar,$^{(2,4,5)}$ Paolo Zanardi$^{(1,2)}$}
\affiliation{$^{(1)}$Department of Physics and Astronomy, $^{(2)}$Center for Quantum Information Science \&
Technology, $^{(3)}$Information Sciences Institute, $^{(4)}$Department of Electrical Engineering, $^{(5)}$Department of Chemistry\\University of Southern California, Los Angeles, California
90089, USA}

\title{Quantum Adiabatic Markovian Master Equations}

\begin{abstract}
We develop from first principles Markovian master equations suited for studying the time evolution of a system evolving adiabatically while coupled weakly to a thermal bath. We derive two sets of
equations in the adiabatic limit, one using the rotating wave (secular) approximation
that results in a master equation in Lindblad form, the other without the
rotating wave approximation but not in Lindblad form. The two equations make markedly different predictions depending on whether or not the Lamb shift is included. Our analysis keeps track of the various time- and energy-scales associated with the various approximations we make, and thus allows for a systematic inclusion of higher order corrections, in particular beyond the adiabatic limit. We use our formalism to study the evolution
of an Ising spin chain in a transverse field and coupled to a thermal bosonic bath, for which we identify four distinct evolution phases. 
While we do not expect this to be a generic feature, in one of these phases dissipation acts to increase the fidelity of the system state relative to the adiabatic ground state.
\end{abstract}

\maketitle

\affiliation{Department of Physics \& Astronomy and Center for Quantum Information Science \&
Technology, University of Southern California, Los Angeles, California
90089, USA}

%
%

%

\section{Introduction}

%
Recent developments in quantum information processing, in particular theoretical \cite{Farhi20042001} and experimental \cite{Dwave} proposals for adiabatic quantum computation, have generated considerable renewed interest in the old topic of quantum adiabatic dynamics \cite{Born:28,Kato:50}. While much work has been done on rigorous formulations of adiabatic approximations for closed quantum systems, e.g., Refs.~\cite{Teufel:book,Jansen:07,PhysRevA.78.052508,boixo_eigenpath_2009,boixo_fast_2010,lidar:102106,Boixo:10,Wiebe:12}, adiabatic evolution in open quantum systems is still a relatively unexplored topic. 
In this regard master equations governing the evolution of a quantum system with a time-dependent Hamiltonian coupled to an external environment or bath are an important tool. 

The study of master equations with time-dependent Hamiltonians is certainly 
not a new topic, going back at least as far as the pioneering work of Davies \& Spohn~\cite{springerlink:10.1007/BF01011696}, who derived an exact master equation for an adiabatic but
infinitesimally weak system-bath interaction. Other, more recent approaches have been attempted, but in each case certain limitations apply. For example, Childs \textit{et al.} used the Lindblad equation with a time-independent Hamiltonian at
each time step as an approximation to the adiabatic evolution of a system with a time-dependent
Hamiltonian \cite{PhysRevA.65.012322}. Sarandy \& Lidar derived a phenomenological adiabatic master equation, based on the idea that in the adiabatic limit the dynamical superoperator can be decomposed in terms of independently evolving Jordan blocks \cite{PhysRevA.71.012331}. This approach is phenomenological in the sense that it does not allow one to derive the various parameters appearing in the final master equation from underlying system and bath Hamiltonians. Approaches based on non-Hermitian effective Hamiltonians, e.g., Refs.~\cite{Garrison1988177,0953-4075-40-2-004,PhysRevA.78.062114}, are necessarily also phenomenological. A rigorous phenomenological master equation derivation was given by Joye \cite{Joye}. Oreshkov \& Calsamiglia connected open system adiabaticity to the theory of noiseless subsystems \cite{PhysRevLett.105.050503}. Thunstr\"om \textit{et al.} derived a master equation from first principles in the physically reasonable joint limit of slow change and weak open system disturbances, but did not elucidate the relative time and energy scales involved in their approximations \cite{PhysRevA.72.022328}. Various authors provided derivations for slow periodic
Hamiltonians \cite{PhysRevA.73.052311,PhysRevLett.105.030401,PhysRevB.84.235140}. Such derivations are valuable and can be made rigorous, but the assumption of periodicity can be excessive, especially in the context of adiabatic quantum computation. Bounds on the validity of the adiabatic approximation for open systems, but without master equations, were presented in Ref.~\cite{PhysRevA.77.042319} (see also Ref.~\cite{lidar:102106}). Various authors derived or studied adiabatic master equations limited to the case of a single qubit, where detailed physical considerations are possible \cite{PhysRevLett.100.060503,PhysRevA.82.062112,PhysRevA.82.022109}. 

Our goal in this work is to derive master equations for adiabatic open system dynamics from first principles, while keeping track of all physical approximations, time- and energy-scales. In this manner we hope to fill a gap in the earlier literature on this topic, and to provide tools allowing for detailed comparisons with experiments satisfying the explicit assumptions behind our approximations. In particular, we derive several Markovian master equations, distinguished by different levels of adiabatic perturbation theory. When we add the rotating wave approximation (sometimes called the secular approximation) we arrive at master equations in Lindblad form \cite{springerlink:10.1007/BF01608499}, for which positivity of the state is guaranteed at all times \cite{Breuer:2002}.  Our formalism allows for the calculation of non-adiabatic corrections, which we also discuss. We apply our master
equations to numerically study the evolution of a transverse field Ising 
chain coupled to a thermal bath, and discuss generic features of the evolution. 

Our basic starting point is the observation that the Markovian and
adiabatic limits are fundamentally compatible, in the sense that a
Markovian bath is ``fast", while an adiabatically evolving system is
``slow". As long as the corresponding timescales are appropriately
matched, it is possible to derive an adiabatic master equation which
is internally consistent. Somewhat more technically, we observe that
in the interaction picture with respect to the unperturbed system and
bath evolutions, where the bath is evolving ``fast" while the system
is evolving ``slowly", and for sufficiently weak system-bath coupling,
it is possible to consistently apply adiabatic perturbation theory to
the time-dependent system operators. This insight allows us to derive
our adiabatic Markovian master equations. Our work is conceptually
similar in its starting point to that of Amin \textit{et al.}
\cite{PhysRevA.80.022303} (see also the Supplementary Information of
Ref.~\cite{Dwave}), but is significantly more general. Some of our
results are also conceptually similar to those found by Kamleitner \&
Shnirman \cite{PhysRevB.84.235140} { by de Vega {\it et al.}
  \cite{Vega:2010fk}}, {and by Yung} \cite{yung_thermal_2008}, but are again more general.  The master equations we derive are a natural generalization of standard time-independent Hamiltonian results \cite{Breuer:2002,blum2010density}, and our master equations reduce to the standard results after freezing the time dependence of the system Hamiltonian.

The structure of this paper is the following. We set the stage in Section~\ref{sec:prelim}, where we define the system-bath model, perform the Born-Markov approximation, and introduce the bath correlation functions and spectral-density. We pay special attention to constraints imposed (via the KMS condition) by baths in thermal equilibrium, and single out the case of the Ohmic oscillator bath. We interrupt the formal development in Section~\ref{sec:time}, where we provide a summary of all the time- and energy-scales appearing in our various approximations, and the inequalities they must mutually satisfy. We then proceed to derive our adiabatic master equations in Section~\ref{sec:derivation}. We proceed in several steps. First, in subsection~\ref{sec:Method1} we find the adiabatic limit of the time-dependent system operators. Next, in subsection~\ref{sec:ME-ad-lim} we find two pairs of master equations in the adiabatic limit, one pair in the interaction picture, the other in the Schr\"{o}dinger picture. The master equations within a given picture are distinguished by whether we apply the adiabatic approximation once or twice. Next, in subsection~\ref{sec:RWA}, we introduce the rotating wave approximation, which allows us to reduce the Schr\"{o}dinger picture master equations into Lindblad form. We conclude the formal development by discussing non-adiabatic corrections to our master equations in subsection~\ref{sec:AdC}. We move on to a detailed numerical study of an example in Section~\ref{sec:example}, of a ferromagnetic chain in a transverse field, coupled to an Ohmic oscillator bath at finite temperature. We apply two of our Schr\"{o}dinger picture master equations, with and without the rotating wave approximation, and show that without inclusion of the Lamb shift term they yield similar predictions. When the Lamb shift is included, however, substantial differences emerge. We discern four distinct phases in the evolution from the transverse field to the ferromagnetic Ising model, which we discuss and analyze. Concluding remarks are presented in Section~\ref{sec:conc}. The paper is supplemented with detailed appendixes where many of the technical details of the derivations and required background are provided, both for ease of flow of the general presentation and for completeness.


\section{Preliminaries}
\label{sec:prelim}
%

\subsection{Model}

We consider a general system-bath Hamiltonian 
%
\begin{equation}
H(t) = H_S(t) + H_B + H_{I} \ ,
\end{equation}
where $H_S(t)$ is the time-dependent, adiabatic system Hamiltonian, $H_B$ is
the bath Hamiltonian, and $H_I$ is the interaction Hamiltonian. Without loss
of generality 
the interaction Hamiltonian can be written in the form: 
\begin{equation}
H_I = {g} \sum_{\alpha} A_{\alpha} \otimes B_{\alpha} \ ,
\end{equation}
where the operators $A_{\alpha}$ and $B_{\alpha}$ are Hermitian {and
dimensionless with unit operator norm, and $g$ is the (weak) system-bath
coupling}. The joint system-bath density matrix $\rho(t)$ satisfies the
Schr\"odinger equation $\dot{\rho} = -i[H,\rho(t)]$, where we have assumed
units such that $\hbar = 1$.

Let $U_{S}(t,t^{\prime })=T_{+}\exp [-i\int_{t^{\prime }}^{t}d\tau \
H_{S}(\tau )]$ and $U_{B}(t,t^{\prime })=\exp (-i(t-t^{\prime })H_{B})$
denote the free system and bath unitary evolution operators ($T_{+}$ denotes
time ordering), and define $U_{0}(t,t^{\prime })=U_{S}(t,t^{\prime })\otimes
U_{B}(t,t^{\prime })$. Likewise, let $U(t,t^{\prime })=T_{+}\exp
[-i\int_{t^{\prime }}^{t}d\tau \ H(\tau )]$ denote the joint Schr\"{o}dinger
picture system-bath unitary evolution operator. We transform to the
interaction picture with respect to $H_{S}(t)$ and $H_{B}$, by defining $%
\tilde{U}(t,0)=U_{0}^{\dagger }(t,0)U(t,0)$, which, together with the
interaction picture density matrix $\tilde{\rho}(t)=U_{0}^{\dagger
}(t,0)\rho (t)U_{0}(t,0)$ satisfies 
\begin{subequations}
\begin{eqnarray}
\frac{d}{dt}\tilde{U}(t,0) &=&-i{\tilde{H}_{I}(t)\tilde{U}(t,0)}\ ,\quad 
\tilde{U}(0,0)=\mathds{1}  \ ,   \\
\frac{d}{dt}\tilde{\rho}(t) &=&-i[\tilde{H}_{I}(t),\tilde{\rho}(t)]\ ,\quad 
\tilde{\rho}(0)={\rho }(0)\ . \label{eq:rhoIP}
\end{eqnarray}
We restrict the use of tilde variables to refer to variables in the
interaction picture. The time-dependent interaction picture Hamiltonian $%
\tilde{H}_{I}(t)$ is related to its Schr\"{o}dinger picture counterpart via: 
\end{subequations}
\begin{equation}
\tilde{H}_{I}(t)=U_{0}^{\dagger }(t,0){H_{I}}U_{0}(t,0)= g \sum_{\alpha
}A_{\alpha }(t)\otimes B_{\alpha }(t)\ ,
\end{equation}
where we have defined the time-dependent system and bath operators: 
%
\begin{subequations}
\begin{eqnarray}
A_{\alpha }(t) &=&U_{S}^{\dagger }(t,0)A_{\alpha }U_{S}(t,0)
\label{eqt:System_Bath_Op} \\
B_{\alpha }(t) &=&U_{B}^{\dagger }(t,0)B_{\alpha }U_{B}(t,0)\ .
\end{eqnarray}
\end{subequations}
We are interested in deriving a master equation for the system-only state,
\bes \label{eqt:6}
\begin{eqnarray}
\tilde{\rho}_{S}(t) &\equiv& \mathrm{Tr}_{B}[\tilde{\rho}(t)]= U_S^\dagger(t,0) \mathrm{Tr}_B( U_B^\dagger (t,0)\rho(t) U_B(t,0)) U_S(t,0)  \\
 &=& U_S^\dagger(t,0) \mathrm{Tr}_B(  U_B(t,0) U_B^\dagger (t,0)\rho(t)) U_S(t,0)  =U_{S}^{\dagger}(t,0)\rho _{S}(t)U_{S}(t,0)\;,
\end{eqnarray}
\ees
where {in the second line we used the fact} that $U_B$ acts only on the bath.

\subsection{Born-Markov approximation}

Writing the formal solution of Eq.~{\eqref{eq:rhoIP}} as 
\begin{equation}
\tilde{\rho}(t)=\tilde{\rho}(0)-i\int_{0}^{t}d\tau \ [\tilde{H}_{I}(\tau ) \ ,%
\tilde{\rho}(\tau )],  \label{eq:formal}
\end{equation}
and substituting this solution back into Eq.~{\eqref{eq:rhoIP}}, we obtain,
after tracing over the bath degrees of freedom, the equation of motion for
the system density matrix $\tilde{\rho}_{S}(t)=\mathrm{Tr}_{B}\tilde{\rho}(t)
$ 
\begin{equation}
\frac{d}{dt}\tilde{\rho}_{S}(t)=-i\mathrm{Tr}_{B}\left[ \tilde{H}_{I}(t) ,%
\tilde{\rho}(0)\right] -\mathrm{Tr}_{B}\left[ \tilde{H}_{I}(t),\int_{0}^{t}d%
\tau \left[ \tilde{H}_{I}(t-\tau ),\tilde{\rho}(t-\tau )\right] \right] \ ,
\label{eq:rho-exact}
\end{equation}%
%
We make the standard Born approximation assumption, that we can decompose
the density matrix as $\tilde{\rho}(t)=\tilde{\rho}_{S}(t)\otimes {\rho%
}_{B}+\chi (t)$ where $\chi (t)$, which expresses correlations between
the system and the bath, is small in an appropriate sense and can hence be
neglected from now on \cite{Breuer:2002,PhysRevA.64.062106,comm1}.
Thus the equation of motion reduces to: 
\begin{equation}
\frac{d}{dt}\tilde{\rho}_{S}(t)={g^{2}}\sum_{\alpha ,\beta
}\int_{0}^{t}d\tau \left( A_{\beta }(t-\tau )\tilde{\rho}_{S}(t-\tau
)A_{\alpha }(t)-A_{\alpha }(t)A_{\beta }(t-\tau )\tilde{\rho}_{S}(t-\tau
)\right) \mathcal{B}_{\alpha \beta }(t,t-\tau )+\mathrm{h.c.}\ ,
\label{eqt:eom1}
\end{equation}
where we defined the two-point correlation functions: 
\begin{equation}
\mathcal{B}_{\alpha \beta }(t,t-\tau )\equiv \langle B_{\alpha }(t)B_{\beta
}(t-\tau )\rangle =\mathrm{Tr}\left[ B_{\alpha }(t)B_{\beta
}(t-\tau ){\rho}_{B}\right] \ .
\end{equation}
In Eq.~\eqref{eqt:eom1}, we have assumed for simplicity that $\langle
B_{\alpha }\rangle _{0}=\mathrm{Tr}\left[ B_{\alpha }(t)\tilde{\rho}%
_{B}(0)\right] =0$, so that the inhomogeneous term in Eq.~%
\eqref{eq:rho-exact} vanishes. Since we assumed that the bath state $\rho
_{B}$ is stationary, the correlation function is homogenous in time: 
\begin{equation}
\mathcal{B}_{\alpha \beta }(t,t-\tau )=\langle B_{\alpha }(t)B_{\beta
}(t-\tau )\rangle =\langle B_{\alpha }(\tau )B_{\beta
}(0)\rangle =\langle B_{\beta
}(0)B_{\alpha }(\tau )\rangle^*=\mathcal{B}_{\alpha \beta }(\tau ,0)\ .
\label{eq:B-prop}
\end{equation}
For notational simplicity we will denote $\mathcal{B}_{\alpha \beta }(\tau
,0)$ by $\mathcal{B}_{\alpha \beta }(\tau )$ when this does not lead to confusion. Let us denote the time scale
over which the two-point correlations of the bath decay by $\tau _{B}$, {e.g., $|\mathcal{B}_{\alpha \beta }(\tau )|\sim \exp (-\tau /\tau _{B})$}. More 
precisely, we shall require throughout that 
\begin{equation}
\int_{0}^{\infty }d\tau \ \tau ^{n}|B_{\alpha \beta }(\tau )|\sim \tau
_{B}^{n+1} \ ,\qquad    {n\in \{0,1,2\}}\ .
\label{eq:corr-decay}
\end{equation}
%

 As we show in Appendix~\ref{app:MarkovTimeScale}, if $\tau _{B}\ll 1/g$, then we can apply the Markov approximation to each of the four
summands in Eq.~\eqref{eqt:eom1}, {i.e., replace $\tilde{\rho}_{S}(t-\tau )$ by $\tilde{\rho}_{S}(t)$}:
\begin{equation}
\int_{0}^{t}d\tau \left( \dots \tilde{\rho}_{S}(t-\tau )\dots \right) 
\mathcal{B}_{\alpha \beta }(\tau )\approx \int_{0}^{\infty }d\tau \left(
\dots \tilde{\rho}_{S}(t)\dots \right) \mathcal{B}_{\alpha \beta }(\tau )+{%
O({\tau _{B}^{3}g^2})}\ ,  
\label{eqt:Markov}
\end{equation}%
where $\left( \dots \right) $ refers to the products of $A_{\alpha }$ and $%
A_{\beta }$ operators in Eq.~\eqref{eqt:eom1}, and where we have also
assumed that {$t\gg \tau_B$ and} the integrand vanishes sufficiently fast for $\tau \gg \tau _{B}
$, so that the upper limit can be taken to infinity. Note that by Eq.~%
\eqref{eq:corr-decay} the integral on the RHS of Eq.~\eqref{eqt:Markov} is
of $O(\tau _{B})$, so that the \emph{relative} magnitude of the two terms is 
$O[{(g\tau _{B})^2}]$. An explicit derivation of the upper bound on the error due
to this approximation can be found in Appendix~\ref{app:MarkovTimeScale}. The resulting Markovian equation cannot resolve the
dynamics over a time scale shorter than $\tau _{B}$. %

\subsection{Correlation functions, {the KMS condition,} and the spectral-density matrix}
\label{subsec:KMS}

In computing the terms appearing in Eq.~\eqref{eqt:Markov} we are faced with
integrals of the form $\int_{0}^{\infty }d\tau A_{\beta }(t-\tau )\tilde{\rho%
}_{S}(t)A_{\alpha }(t)\mathcal{B}_{\alpha \beta }(\tau )$ and $%
\int_{0}^{\infty }d\tau A_{\alpha }(t)A_{\beta }(t-\tau )\tilde{\rho}_{S}(t)%
\mathcal{B}_{\alpha \beta }(\tau )$. Our goal is to express these integrals
in terms of the \emph{spectral-density matrix}%
\begin{equation} \label{eqt:SpectralDensity}
\Gamma _{\alpha \beta }(\omega )\equiv \int_{0}^{\infty }d\tau e^{i\omega
\tau }\mathcal{B}_{\alpha \beta }(\tau )\ ,
\end{equation}
the standard quantity in master equations. {It is convenient to replace the one-sided Fourier transform in the spectral-density matrix by a complete Fourier transform. Thus we split it into a sum of Hermitian matrices,}
\begin{equation} 
\label{eqt:Gamma}
\Gamma_{\alpha \beta}(\omega) = \frac{1}{2} \gamma_{\alpha \beta} (\omega) + i S_{\alpha \beta}(\omega)  \ ,
\end{equation}
where we show in appendix \ref{app:Gamma} that $\gamma$ and $S$ are given by
\bes
\label{eq:gS}
\bea
\label{eq:gamma_def}
\gamma_{\alpha \beta}(\omega) &=& \int_{-\infty}^\infty d \tau e^{i \omega
\tau} \mathcal{B}_{\alpha \beta}(\tau ) = \gamma^*_{\beta \alpha}(\omega) {\ , } \\
S_{\alpha\beta}(\omega) &=& \int_{-\infty}^{\infty} \frac{d\omega^{\prime }}{2\pi}
\gamma_{\alpha \beta}(\omega^{\prime }) \mathcal{P}\left( \frac{1}{\omega -
\omega^{\prime }} \right) = S^*_{\beta \alpha}(\omega) \ .
\eea
\ees
If we assume not only that the bath state is stationary, but that it is also in thermal equilibrium at inverse temperature $\beta$, i.e., $\rho_B=e^{-\beta H_B}/\mathcal{Z}$, then it follows that the correlation function satisfies the KMS (Kubo-Martin-Schwinger) condition \cite{Breuer:2002}
\beq
\label{eq:KMSt}
\langle B_a(\tau)B_b(0)\rangle = \langle B_b(0)B_a(\tau+i\beta)\rangle  \ .
\eeq
{If in addition the correlation function is analytic in the strip between $\tau=-i\b$ and $\tau=0$, then it follows that the Fourier transform of the bath correlation function} satisfies the detailed balance condition
\beq
\label{eq:KMS}
\gamma_{ab}(-\omega) = e^{-\beta\omega}\gamma_{ba}(\omega)\ .
\eeq
For a proof see Appendix~\ref{app:KMS}.

{It is important to note that the KMS detailed balance condition imposes a restriction on the decay of the correlation function. To see this, note first that $\left\vert \mathcal{B}_{ab}(-\tau )\right\vert
=\left\vert \mathrm{Tr}[\rho _{B}U_{B}(\tau )B_{a}U_{B}^{\dag }(\tau
)B_{b}]\right\vert =\left\vert \mathrm{Tr}[B_{a}U_{B}^{\dag }(\tau
)B_{b}U_{B}(\tau )\rho _{B}]\right\vert =\left\vert \langle
B_{a}(0)B_{b}(\tau )\rangle \right\vert =\left\vert \langle B_{b}(\tau
)B_{a}(0)\rangle \right\vert =\left\vert \mathcal{B}_{ba}(\tau )\right\vert $, where we used Eq.~\eqref{eq:B-prop}. Now assume that Eq.~\eqref{eq:corr-decay} would have to hold for all $n$. It would follow that
\begin{eqnarray}
\left\vert  \frac{d^{n}}{d\omega ^{n}}\gamma _{ab}(\omega )\right\vert
_{\omega =0}  &=&\left\vert \int_{-\infty }^{\infty }\left. \tau
^{n}e^{i\omega \tau }\mathcal{B}_{ab}(\tau )d\tau \right\vert _{\omega
=0}\right\vert =\left\vert \int_{-\infty }^{\infty }\tau ^{n}\mathcal{B}%
_{ab}(\tau )d\tau \right\vert \leq \int_{0}^{\infty }\tau ^{n}\left(
\left\vert \mathcal{B}_{ba}(\tau )\right\vert +\left\vert \mathcal{B}%
_{ab}(\tau )\right\vert \right) d\tau   \nonumber \\
&\sim &2\tau _{B}^{n+1}\quad \forall n\in \{0,1,\ldots \},
\label{eq:dgamma-n}
\end{eqnarray}%
Thus all derivatives of $\gamma _{ab}(\omega )$ would have to be finite at 
$\omega =0$. }

On the other hand, it follows from the KMS condition \eqref{eq:KMS} that 
\begin{eqnarray}
\left[ \frac{d}{d(-\omega) }\gamma _{ab}(-\omega ) \right]_{\omega \geq 0}=
 \left. \left[ \b e^{-\b \omega}\gamma_{ba}(\omega) - e^{-\b \omega}\frac{d}{d\omega }\gamma _{ba}(\omega)\right]\right\vert _{\omega \geq 0} \ , 
\end{eqnarray}
so that in the limit as $\omega$ approaches zero from below or above,
\begin{eqnarray}
\gamma_{ab}'(0_-) = \beta \gamma_{ba}(0) - \gamma_{ba}'(0_+) \ .
\label{eq:KMS'}
\end{eqnarray}
This shows that in principle $\gamma'_{aa}(\omega)$ may be discontinuous at $\omega=0$. Indeed, the continuity condition 
$ \gamma_{aa}'(0_-) =  \gamma_{aa}'(0_+)$ 
implies, 
from the KMS condition recast as Eq.~\eqref{eq:KMS'}, that
\begin{equation} \label{eqt:deriv}
2 \gamma'_{aa}(0) = \beta \gamma_{aa}(0) \ . 
\end{equation}
This conclusion can be extended to the entire $\gamma$ matrix by diagonalizing it first. When Eq.~\eqref{eqt:deriv} is not satisfied $\gamma''_{aa}(0)$ diverges, so that Eq.~\eqref{eq:dgamma-n} does not hold except for $n\in\{0,1\}$. A simple example of this is $\gamma _{ab}(\omega )=c>0$ (constant) for $\omega \geq 0$.  Another example is a super-Ohmic spectral-density $\gamma_{ab}(\omega) = \omega^2/(1-e^{-\beta\omega})$ for $\omega\geq 0$ and $\gamma_{ab}(\omega) = \omega^2/(e^{-\beta\omega}-1)$ for $\omega\leq 0$. Both examples violate Eq.~\eqref{eq:dgamma-n} for $n\geq 2$. 
Note, moreover, that when this happens, it follows from Eq.~\eqref{eq:dgamma-n} that $\int_{0}^{\infty }\tau ^{2}\left(\left\vert \mathcal{B}_{ab}(-\tau )\right\vert +\left\vert \mathcal{B}%
_{ab}(\tau )\right\vert \right) d\tau$ is divergent, so that we must conclude that $|\mathcal{B}_{ab}(\tau )| \gtrsim 1/\tau^3$ for sufficiently large $|\tau|$, meaning that the correlation function has a power-law tail and, in particular, cannot be exponentially decaying.

On the other hand, Eq.~\eqref{eqt:deriv} tells us that continuity and a lack of divergence at $\omega=0$ require 
$\gamma'_{aa}(0)$ to satisfy a condition which prohibits it from being
arbitrary. 
This is indeed the case for the Ohmic oscillator bath discussed in subsection~\ref{sec:Model}, which satisfies Eq.~\eqref{eqt:deriv} with finite $\gamma_{aa}(0)$. For this case the bath correlation function is exponentially decaying assuming the oscillator bath has an infinite cutoff. {However, as we show in subsection~\ref{subsec:cor-analysis} the Ohmic bath transitions from exponential decay to a power-law tail for any finite value of the cutoff, at some finite transition time $\tau_{\mathrm{tr}}$. In this case we find $|B_{\alpha \beta }(\tau )|\sim (\tau/\tau_M)^{-2}$, where $\tau_M$ is a time-scale associated with the onset of non-Markovian effects, and hence we have to relax Eq.~\eqref{eq:corr-decay}, and, replace it with 
\bes
\label{eq:corr-decay2}
\begin{align}
&\int_{0}^{\tau_{\mathrm{tr}}}d\tau \ \tau ^{n}|B_{\alpha \beta }(\tau )| \sim \tau
_{B}^{n+1} \ ,\qquad    {n\in \{0,1,\dots\}} \\
&\int_{\tau_{\mathrm{tr}}}^{\infty}d\tau \ |B_{\alpha \beta }(\tau )| \sim  \int_{\tau_{\mathrm{tr}}}^{\infty}d\tau \ (\tau/\tau_M)^{-2} = \tau_M^2/\tau_{\mathrm{tr}}\ .
\end{align}
\ees
}

\section{Timescales}
\label{sec:time}

%
In this subsection we summarize, for convenience, the relations between the
different timescales which shall arise in our derivation. The total
evolution time is denoted $t_f$ and the minimum ground state energy gap of $%
H_S$ is $\Delta$, i.e., 
\begin{equation}
\Delta \equiv \min_{t\in[0,t_f]} \varepsilon_1(t) - \varepsilon_0(t),
\end{equation}
where $\varepsilon_0(t)$ and $\varepsilon_1(t)$ are the ground and first
excited state energies of $H_S(t)$. We shall arrive at master equations of
the following general form: 
\begin{equation}  \label{eq:L}
\dot{\rho}_S(t) = [\mathcal{L}_{\text{uni}}(t)+\mathcal{L}_{\text{diss}}^{%
\text{ad}}(t)+\mathcal{L}_{\text{diss}}^{\text{non-ad}}(t)]\rho_S(t) \ .
\end{equation}
Here $\mathcal{L}_{\text{uni}} = -i[H_S(t) + H_{\textrm{LS}},\cdot]$ is the unitary
evolution superoperator including the Lamb shift correction, $\mathcal{L}_{\text{diss}}^{\text{ad}}$ is the dissipative
superoperator in the fully adiabatic limit, and $\mathcal{L}_{\text{diss}}^{%
\text{non-ad}}$ is the dissipative superoperator with leading order
non-adiabatic corrections.

Let
\begin{equation}  \label{eq:h}
h\equiv\max_{s\in[0,1];a,b} |\langle \varepsilon_a (s) | \partial_s H_S(s) |
\varepsilon_b (s) \rangle | \ ,
\end{equation}
where $s\equiv t/t_f$ is the dimensionless time. To ensure that $\mathcal{L}_{\text{uni}}$ generates adiabatic evolution to
leading order we shall require the standard adiabatic condition 
\begin{equation}  
\label{eq:ad-cond}
\frac{h}{\Delta^2 t_f}\ll1 \ .
\end{equation}

In order to derive Eq.~\eqref{eq:L}, the three superoperators are ordered in
perturbation theory, in the sense that 
\begin{equation}  \label{eq:L-order}
\|\mathcal{L}_{\text{uni}}\| \gg \|\mathcal{L}_{\text{diss}}^{\text{ad}}\|
\gg \|\mathcal{L}_{\text{diss}}^{\text{non-ad}}\| \ ,
\end{equation}
where the norm could be chosen as the supoperator norm, i.e., the largest
singular value (see Appendix~\ref{app:norms}). {We may then assume that 
\begin{equation}
\|\mathcal{L}_{\text{uni}}\| \gg \Delta \gg \|\mathcal{L}_{\text{diss}}^{\text{ad}}\| \ .
\end{equation}
}%
Combining the $O(\tau _{B})$ due to the integral on the RHS of Eq.~\eqref{eqt:Markov} (as already remarked there) with the $g^2$ prefactor from Eq.~\eqref{eqt:eom1}, we have 
\begin{equation}
\|\mathcal{L}_{\text{diss}}^{\text{ad}}\| \sim g^2 \tau_B \ .
\end{equation}
To ensure the first inequality in
Eq.~\eqref{eq:L-order} we thus require 
\begin{equation}  \label{eq:cond2}
\frac{g^2\tau_B}{\Delta} \ll 1 \ . 
\end{equation}
The non-adiabatic correction is of order $\frac{h}{\Delta^2 t_f}$, and when it appears in $\mathcal{L}_{\text{diss}}^{\text{non-ad}}$ it is multiplied by the same factor as $\mathcal{L}_{\text{diss}}^{\text{ad}}$, i.e., we have 
\begin{equation}
\|\mathcal{L}_{\text{diss}}^{\text{non-ad}}\| \sim g^2 \tau_B \frac{h}{\Delta^2 t_f} \ .
\end{equation}
To ensure the second inequality in Eq.~%
\eqref{eq:L-order} thus amounts to the adiabatic condition, Eq.~%
\eqref{eq:ad-cond}. All this is added to the condition 
\begin{equation}  \label{eq:Markov-cond}
g \tau_B \ll 1 \ ,
\end{equation}
for the validity of the Markovian approximation, mentioned in the context of
Eq.~\eqref{eqt:Markov}, and justified rigorously in Appendix~\ref{app:MarkovTimeScale}.

There is one additional, {independent} time scale we have to concern
ourselves with. This is the time scale associated with changes in the
instantaneous energy eigenbasis relative to $\tau_B$. If we require the
change in the basis to be small on the time scale of the bath $\tau_B$, 
we must have: 
\begin{equation}
\frac{h\tau_B^2}{t_f} \ll 1 \ .  
\label{eq:ts-4}
\end{equation}
We justify this claim in subsection~\ref{sec:Method1} and Appendix~\ref{app:ShortTime}.

Note that the adiabatic condition, Eq.~\eqref{eq:ad-cond}, implies $\frac{h
\tau_B}{\Delta t_f} \ll \Delta \tau_B$, while Eq.~\eqref{eq:ts-4} can be written as $\frac{h\tau_B}{\Delta t_f} \ll \frac{1}{\Delta\tau_B}$.
Putting this together thus yields 
\begin{equation}
\frac{h \tau_B}{\Delta t_f} \ll \min(\Delta \tau_B,\frac{1}{\Delta\tau_B}) \ .
\label{eq:strongad}
\end{equation}
Our other inequalities [Eqs.~\eqref{eq:cond2}, \eqref{eq:Markov-cond}] can
be summarized as  
\begin{equation}
\label{eq:gtau>}
g \tau_B \ll \min(1,{\Delta/g}) \ .
\end{equation}


\section{Derivation of Adiabatic Master Equations}
\label{sec:derivation}

\subsection{Adiabatic limit of the time-dependent system operators}

\label{sec:Method1} 
%
Let us first work in the strict adiabatic limit. We will discuss
non-adiabatic corrections in subsection~\ref{sec:AdC}. We denote the
instantaneous eigenbasis of $H_{S}(t)$ by $\{\ket{\varepsilon_a(t)}\}$, with
corresponding real eigenvalues (energies) $\{{\varepsilon _{a}(t)}\}$, i.e., 
$H_{S}(t)\ket{\varepsilon_a(t)}={\varepsilon _{a}(t)}\ket{\varepsilon_a(t)}$, and Bohr frequencies $\omega _{ba}(t)\equiv \varepsilon
_{b}(t)-\varepsilon _{a}(t).$ As shown in Appendix~\ref{app:AdC} we can then
write the system time evolution operator as: 
\bes
\begin{eqnarray}
U_{S}(t,t^{\prime }) &=&U_{S}^{\text{ad}}(t,t^{\prime })+O\left( \frac{h}{%
\Delta ^{2}t_{f}}\right)  {\ , } \label{eqt:AdTimeEvol} \\
U_{S}^{\text{ad}}(t,t^{\prime }) &=&\sum_{a}|\varepsilon _{a}(t)\rangle
\langle \varepsilon _{a}(t^{\prime })|e^{-i\mu _{a}(t,t^{\prime })} { \ ,}
\label{eq:U_S^ad}
\end{eqnarray}
\ees
where $U_{S}^{\text{ad}}(t,t^{\prime })$ is the ``ideal"
adiabatic evolution operator. It represents a transformation of
instantaneous eigenstate $|\varepsilon _{a}(t^{\prime })\rangle $ into the
later eigenstate $|\varepsilon _{a}(t)\rangle $, along with a phase 
\beq
\mu _{a}(t,t^{\prime })= \int_{t^{\prime }}^{t}d\tau \left[\varepsilon
_{a}(\tau )-\phi_a(\tau)\right]\ ,
\eeq 
where $\phi_a(t) = i\langle \varepsilon _{a}(t)|\dot{\varepsilon}_{a}(t)\rangle$ is the Berry connection. The correction term $%
O\left( \frac{h}{\Delta ^{2}t_{f}}\right) $ is derived in Appendix~\ref{app:adiabatic-time}.

To achieve our goal of arriving at a master equation expressed in terms of the spectral-density matrix, our basic
strategy is to replace the system operator $A_{\beta }(t-\tau )=U_{S}^{\dag
}(t-\tau ,0)A_{\beta }U_{S}(t-\tau ,0)$ with an appropriate adiabatic
approximation, which will allow us to take this operator outside of the
integral. To see how, note first that 
\begin{equation}
U_{S}(t-\tau ,0)=U_{S}(t-\tau ,t)U_{S}(t,0)=U_{S}^{\dag }(t,t-\tau
)U_{S}(t,0) \ .
\end{equation}
We now make two approximations:\ first, as per Eq.~\eqref{eqt:AdTimeEvol} we
replace $U_{S}(t,0)$ by $U_{S}^{\text{ad}}(t,0)$; second, we replace $%
U_{S}^{\dag }(t,t-\tau )$ by $e^{i\tau H_{S}(t)}$, an approximation
justified by the appearance of the short-lived bath correlation function $%
\mathcal{B}_{\alpha \beta }(\tau )$ inside the integrals we are concerned
with. Thus, we write
\begin{equation}
U_{S}(t-\tau ,0)=e^{i\tau H_{S}(t)}U_{S}^{\text{ad}}(t,0)+\Theta (t,\tau ) \ ,
\label{eq:U-approx}
\end{equation}
and find the bound on the error due to dropping $\Theta (t,\tau )$ in Appendix~\ref{app:ShortTime}. Let  
\bes
\begin{eqnarray}
\mu _{ba}(t,t^{\prime }) &\equiv &\mu _{b}(t,t^{\prime })-\mu
_{a}(t,t^{\prime }) \ , \\
\Pi _{ab}(t) &\equiv &|\varepsilon _{a}(t)\rangle \langle \varepsilon
_{b}(t)| \ .
\end{eqnarray}%
\ees
Neglecting the operator-valued correction term $\Theta
(t,\tau )$ entirely, we have, upon substituting Eq.~\eqref{eq:U_S^ad} and
using $e^{i\tau H_{S}(t)}|\varepsilon _{a}(t)\rangle =e^{i\tau \varepsilon
_{a}(t)}|\varepsilon _{a}(t)\rangle $, that   
\bes
\begin{align}
&\int_{0}^{\infty }d\tau A_{\beta }(t-\tau )\tilde{\rho}_{S}(t)A_{\alpha }(t)%
\mathcal{B}_{\alpha \beta }(\tau ) \\
\label{eq:33b}
&\quad \approx\int_{0}^{\infty }d\tau U_{S}^{\text{ad}\dag }(t,0)e^{-i\tau H_{S}(t)}A_{\beta }e^{i\tau H_{S}(t)}U_{S}^{\text{ad}}(t,0)\tilde{\rho}_{S}(t)A_{\alpha }(t)\mathcal{B}_{\alpha \beta }(\tau
)  \\
&\quad =\int_{0}^{\infty }d\tau \sum_{ab}e^{-i\mu _{ba}(t,0)}|\varepsilon
_{a}(0)\rangle \langle \varepsilon _{a}(t)|e^{-i\tau H_{S}(t)}A_{\beta
}e^{i\tau H_{S}(t)}|\varepsilon _{b}(t)\rangle \langle \varepsilon _{b}(0)|%
\tilde{\rho}_{S}(t)A_{\alpha }(t)\mathcal{B}_{\alpha \beta }(\tau )  
\label{eq:33c} \\
&\quad =\sum_{ab}e^{-i\mu _{ba}(t,0)}|\varepsilon _{a}(0)\rangle \langle
\varepsilon _{a}(t)|A_{\beta }|\varepsilon _{b}(t)\rangle \langle
\varepsilon _{b}(0)|\tilde{\rho}_{S}(t)A_{\alpha }(t)\int_{0}^{\infty }d\tau
e^{i\tau \lbrack \varepsilon _{b}(t)-\varepsilon _{a}(t)]}\mathcal{B}%
_{\alpha \beta }(\tau ) \ ,
\end{align}%
\ees
where the approximation in \eqref{eq:33b} is shown in Appendix~\ref{app:ShortTime} to be {$O[\tau_B\min\{1,\frac{h}{\Delta^2 t_f} + \frac{\tau_B^2 h}{t_f}\}]$. The first term, $\frac{h}{\Delta^2 t_f}$, is the smallness parameter of the adiabatic approximation, which we have already assumed to be small. The second term, $\frac{\tau_B^2 h}{t_f}$ [mentioned in Eq.~\eqref{eq:ts-4}], is new, and its smallness is associated with changes in the instantaneous energy eigenbasis relative to $\tau_B$. We are interested in the regime where both terms are small.} 

Thus, to the same level of approximation
\begin{equation}
\int_{0}^{\infty }d\tau A_{\beta }(t-\tau )\tilde{\rho}_{S}(t)A_{\alpha }(t)%
\mathcal{B}_{\alpha \beta }(\tau )\approx \sum_{ab}e^{-i\mu _{ba}(t,0)}A_{\beta
ab}(t)\Pi _{ab}(0)\tilde{\rho}_{S}(t)A_{\alpha }(t)\Gamma _{\alpha \beta
}(\omega _{ba}(t))\ ,  \label{eq:ArA}
\end{equation}
where 
\begin{equation}
A_{\alpha ab}(t)\equiv \langle \varepsilon _{a}(t)|A_{\alpha }|\varepsilon
_{b}(t)\rangle = A^*_{\alpha ba}(t) \ ,  \label{eqt:A_define}
\end{equation}
and $\Gamma _{\alpha \beta}(\omega _{ba}(t))$ is the spectral-density matrix defined in Eq.~\eqref{eqt:SpectralDensity}.
Similarly, we have for the other term
\begin{equation}
\int_{0}^{\infty }d\tau A_{\alpha }(t)A_{\beta }(t-\tau )\tilde{\rho}_{S}(t)%
\mathcal{B}_{\alpha \beta }(\tau )\approx\sum_{ab}e^{-i\mu _{ba}(t,0)}A_{\beta
ab}(t)A_{\alpha }(t)\Pi _{ab}(0)\tilde{\rho}_{S}(t)\Gamma _{\alpha \beta
}(\omega _{ba}(t)) .  \label{eq:AAr}
\end{equation}

\subsection{Master equations in the adiabatic limit}
\label{sec:ME-ad-lim}

We are now ready to put everything together. Starting from the Born-Markov
master equation constructed from Eqs.~\eqref{eqt:eom1} and \eqref{eqt:Markov}%
, and using the approximations \eqref{eq:ArA} and~\eqref{eq:AAr}, we arrive
at the following \emph{one-sided adiabatic interaction picture master
equation}: 
\begin{equation}
\frac{d}{dt}\tilde{\rho}_{S}(t)=g^{2}\sum_{ab}e^{-i\mu
_{ba}(t,0)}\sum_{\alpha \beta }\Gamma _{\alpha \beta }(\omega
_{ba}(t))A_{\beta ab}(t)\left[ \Pi _{ab}(0)\tilde{\rho}_{S}(t),A_{\alpha }(t)%
\right] +\mathrm{h.c.}\ . 
\label{eqt:newEOM}
\end{equation}%

Since we used an adiabatic approximation for $A_{\beta
}(t-\tau )$, it makes sense to do the same for $A_{\alpha }(t)$, i.e., to
replace the latter with $U_{S}^{\text{ad}\dag }(t,0)A_{\alpha }U_{S}^{\text{%
ad}}(t,0)$. If this is done, we obtain the \emph{double-sided adiabatic
interaction picture master equation}%
\begin{equation}
\frac{d}{dt}\tilde{\rho}_{S}(t)=g^{2}\sum_{abcd}e^{-i\left[ \mu
_{dc}(t,0)+\mu _{ba}(t,0)\right] }\sum_{\alpha \beta }\Gamma _{\alpha \beta
}(\omega _{ba}(t))A_{\alpha cd}(t)A_{\beta ab}(t)\left[ \Pi _{ab}(0)\tilde{%
\rho}_{S}(t),\Pi _{cd}(0)\right] +\mathrm{h.c.}\ .
\label{eq:2IME}
\end{equation}

It is convenient to transform back into the Schr\"{o}dinger picture. Using $%
\tilde{\rho}_{S}(t)=U_{S}^{\dag }(t,0)\rho _{S}(t)U_{S}(t,0)$ [Eq.~%
\eqref{eqt:6}] implies that $\frac{d}{dt}\tilde{\rho}_{S}(t)=U_{S}^{\dag
}(t,0)\left( i[H_{S}(t),\rho _{S}(t)]+\frac{d}{dt}{\rho }_{S}(t)\right)
U_{S}(t,0)$. Hence, using Eq.~\eqref{eqt:newEOM}, we find the \emph{%
one-sided adiabatic Schr\"{o}dinger picture master equation}  
\begin{equation}
\frac{d}{dt}{\rho }_{S}(t)=-i[H_{S}(t),\rho
_{S}(t)]+g^{2}\sum_{ab}\sum_{\alpha \beta }\Gamma _{\alpha \beta }(\omega
_{ba}(t))\left[ L_{ab,\beta }^{\prime }(t)\rho _{S}(t),A_{\alpha }\right] +%
\mathrm{h.c.}\ ,
\label{eq:1SME}
\end{equation}%
where%
\begin{equation}
L_{ab,\beta }^{\prime }(t)=e^{-i\mu _{ba}(t,0)}A_{\beta ab}(t)U_{S}(t,0)\Pi
_{ab}(0)U_{S}^{\dag }(t,0).
\end{equation}%
This form of the master equation 
has not appeared in previous studies of adiabatic master equations.
If we again use the adiabatic approximation for $U_{S}(t,0)$, i.e., replace $%
U_{S}(t,0)\Pi _{ab}(0)U_{S}^{\dag }(t,0)$ by $U_{S}^{\text{ad}\dag }(t,0)\Pi
_{ab}(0)U_{S}^{\text{ad}}(t,0)$, we obtain the \emph{double-sided adiabatic
Schr\"{o}dinger picture master equation}       
\begin{equation}
\frac{d}{dt}{\rho }_{S}(t)=-i\left[ H_{S}(t),\rho _{S}(t)\right]
+g^{2}\sum_{\alpha \beta }\sum_{ab}\Gamma _{\alpha \beta }(\omega _{ba}(t))%
\left[ L_{ab,\beta }(t)\rho _{S}(t),A_{\alpha }\right] +\mathrm{h.c.}\ ,
\label{eqt:NRWA}
\end{equation}%
where 
\begin{equation}
L_{ab,\alpha }(t)\equiv A_{\alpha ab}(t)\ket{\varepsilon_a(t)}\!%
\bra{\varepsilon_b(t)} = L^\dag_{ba,\alpha }(t)\ . 
 \label{eqt:LindbladL}
\end{equation}%
Comparing to Eq.~\eqref{eq:L}, the first term in Eqs.~\eqref{eq:1SME} and \eqref{eqt:NRWA} is $%
\mathcal{L}_{\text{uni}}(t)$, while the second is $\mathcal{L}_{\text{diss}%
}^{\text{ad}}(t)$. 

The master equations we have found so far are not in Lindblad form, and hence do not guarantee the preservation of positivity of $\rho_S$. We thus introduce an additional approximation, which will transform the master equations into completely positive form.

\subsection{Master equation in the adiabatic limit with rotating wave
approximation: Lindblad form}
\label{sec:RWA}

In order to arrive at a master equation in Lindblad form, we can perform a secular, or
rotating wave approximation (RWA).
To do so, let us revisit the $t\rightarrow \infty$ limit taken in the Markov approximation in Eq.~\eqref{eqt:Markov}. Supposing we do not take this limit quite yet, we can follow the same arguments leading to Eq.~\eqref{eq:33c}, which we rewrite, along with the adiabatic approximation $A_\alpha(t)\approx U_{S}^{\text{ad}\dag }(t,0)A_\alpha U_{S}^{\text{ad}}(t,0)$. This yields
\bes
\begin{align}
&\int_{0}^{t}d\tau A_{\beta }(t-\tau )\tilde{\rho}_{S}(t)A_{\alpha }(t)%
\mathcal{B}_{\alpha \beta }(\tau )\approx  \\
&\quad \int_{0}^{t}d\tau \sum_{abcd}e^{-i\left[\mu _{ba}(t,0)+\mu_{dc}(t,0)\right]}|\varepsilon
_{a}(0)\rangle \langle \varepsilon _{a}(t)|A_{\beta
}|\varepsilon _{b}(t)\rangle \langle \varepsilon _{b}(0)|%
\tilde{\rho}_{S}(t) \ketbra{\varepsilon_c(0)}{\varepsilon_c(t)}A_{\alpha }\ketbra{\varepsilon_d(t)}{\varepsilon_d(0)}e^{i\tau\omega_{ba}(t)}\mathcal{B}_{\alpha \beta }(\tau )  \ .
\label{eq:43b}
\end{align}
\ees
We note that $\mu
_{dc}(t,0)+\mu _{ba}(t,0) = \int_0^t d\tau \left[\omega_{dc}(\tau)+\omega_{ba}(\tau)-(\phi_d(\tau)-\phi_c(\tau))+(\phi_b(\tau)-\phi_a(\tau))\right]$. One can now make the argument that when the $t\rightarrow \infty$ limit is taken, terms for which the integrand vanishes will dominate, thus enforcing the ``energy conservation" condition {$\omega_{ba} = -\omega_{dc}$}.  This is a similar rotating wave approximation as made in the standard time-independent treatment, although here, the approximation of phase cancellation is made along the entire time evolution of the instantaneous energy eigenstates.  Clearly, in light of the appearance of other terms involving $t$ in Eq.~\eqref{eq:43b}, this argument is far from rigorous.
Nevertheless, we proceed from 
Eq.~\eqref{eq:43b} to write, in the $t\rightarrow \infty$ limit,
\begin{align}
&\int_{0}^{t}d\tau A_{\beta }(t-\tau )\tilde{\rho}_{S}(t)A_{\alpha }(t)%
\mathcal{B}_{\alpha \beta }(\tau )\approx   
\sum_{\omega}A_{\beta, \omega}(t)A_{\alpha, \omega}(t)\Pi_{\omega}(0)
\tilde{\rho}_{S}(t)\Pi_{\omega}(0) \Gamma_{\alpha\beta}(\omega)  \ ,
\label{eq:46}
\end{align}

where we have defined a new index $\omega$ such that:
\beq
A_{\alpha, \omega}(t) = \sum_{\varepsilon_b(t) - \varepsilon_a(t) = \omega} \langle \varepsilon_a(t) | A_{\alpha} | \varepsilon_b(t) \rangle \ , \quad \Pi_{\omega}(0) = \sum_{\varepsilon_b(t) - \varepsilon_a(t) = \omega}  | \varepsilon_a(0) \rangle  \langle \varepsilon_b(0)| \ .
\eeq
Note that the set of $\{ \omega \}$'s involved in the sum $\sum_{\omega}$ is evolving in time since it corresponds to differences of the instantaneous energy eigenvalues, but we suppress the time dependence for notational brevity.
We show in Appendix~\ref{app:RWA} how, by performing a transformation back to the Schr\"{o}dinger picture, along with a double-sided adiabatic approximation, we arrive from Eq.~\eqref{eq:46} at 
the \emph{Schr\"{o}dinger picture adiabatic master equation in Lindblad form}:
\beq
\label{eqt:RWA}
\dot{\rho}_S(t) = - i \left[ H_S(t) + H_{\textrm{LS}}(t), \rho_S(t) \right] + 
\sum_{\alpha \beta} \sum_{\omega} \gamma_{\alpha \beta}(\omega) \left[
L_{\omega, \beta}(t) \rho_S(t) L^\dagger_{\omega, \alpha}(t)  - \frac{1}{2} \left\{ L^\dagger_{\omega, \alpha}(t) L_{\omega,\beta}(t), \rho_S(t) \right\}\right]   \ , 
\eeq
where the Hermitian Lamb shift term is
\begin{eqnarray}  
\label{eqt:H_LS}
H_{\textrm{LS}}(t) =\sum_{\alpha \beta} \sum_{\omega} L^\dagger_{\omega, \alpha}(t) L_{\omega
, \beta}(t) S_{\alpha \beta}(\omega)  \ , 
\end{eqnarray}
and we have defined
\beq
L_{\omega,\alpha}(t) \equiv  \sum_{\varepsilon_b(t) - \varepsilon_a(t) =\omega} L_{ab,\alpha}(t) \ .
\eeq
Since the bath correlations are of positive type, it follows from Bochner's
theorem that the matrix $\gamma$---the Fourier transform of the bath correlation functions---is also
positive \cite{Breuer:2002}. Therefore, this Lindblad form for our master equation guarantees
the positivity of the density matrix.

We emphasize that Eqs.~\eqref{eq:1SME}, \eqref{eqt:NRWA}, and \eqref{eqt:RWA} all generalize both the standard Redfield and Lindblad time-independent Hamiltonian results to the case of a time-dependent Hamiltonian in the adiabatic limit.\footnote{See for example Eq.~(3.143) in Ref.~\cite{Breuer:2002} and Eq.~(8.1.33) in Ref.~\cite{blum2010density}.}   The time-independent result can easily be recovered by simply freezing the time dependence of the Hamiltonian, energy eigenvalues and eigenvectors.  To see this explicitly, let us restrict ourselves to the Lindblad case.  

The energy eigenvalues and eigenvectors are time independent in this case, so we can replace our time dependent Lindblad operators with time independent ones, and the $\omega$ index no longer varies with time:
\beq
 L_{\omega,\alpha}(t) \to  L_{\omega, \alpha} = \sum_{\varepsilon_b - \varepsilon_a = \omega} | \varepsilon_a \rangle \langle \varepsilon_a | A_{\alpha} | \varepsilon_b \rangle \langle \varepsilon_b| \ .
\eeq
The resulting equation becomes:
\begin{equation}
\dot{\rho}_S(t) = -i \left[H_S + H_{\mathrm{LS}}, \rho(t) \right] + \sum_{\alpha \beta} \sum_{\omega} \gamma_{\alpha \beta} (\omega) \left(L_{\omega, \beta} \rho_S(t) L_{\omega, \alpha}^{\dagger} - \frac{1}{2} \left\{L_{\omega, \alpha}^{\dagger} L_{\omega, \beta}, \rho_S(t) \right\} \right) \ ,
\end{equation}
which is the standard form for the time-independent Lindblad master equation.  This should make evident the physical meaning of our derivation.  We have systematically generalized the time-independent result such that the Lindblad operators now rotate with the (adiabatically) changing energy eigenstates, which makes them time-dependent.  This is a non-trivial difference, as it encodes an important physical effect: \emph{the dissipation/decoherence of the system occurs in the instantaneous energy eigenbasis}.

Equations~\eqref{eqt:NRWA} and \eqref{eqt:RWA} are the two master equations we use for numerical simulations presented later in this
paper. {We note that Eq.~\eqref{eqt:RWA} appears similar to the Markovian adiabatic master equation found in Ref.~\cite{PhysRevB.84.235140}, but is more general and did not require the assumption of periodic driving.}

\subsection{Non-Adiabatic Corrections to the Master Equations}

\label{sec:AdC} 
%
So far, we assumed the adiabatic limit of evolution of the system [see Eq.~%
\eqref{eqt:AdTimeEvol}]. The adiabatic perturbation theory we review in Appendix~\ref{app:AdC} allows us to compute systematic non-adiabatic corrections to the master equations we have derived. This perturbation theory is essentially an expansion in powers of $1/t_f$, and we 
rederive in Appendix~\ref{app:AdC} the well known result \cite{Teufel:book} that to first order 
\beq
U_S(t,t') = U^{\textrm{ad}}_S(t,t')\left[\ident+ V_1(t,t')\right] {\ , }
\eeq
where
\begin{eqnarray}
V_1(t,t') =- \sum_{a \neq b} |\varepsilon_a(t') \rangle \langle \varepsilon_b(t') |\int_{t'}^t d \tau e^{- i \mu_{ba}(\tau,t')}  \langle \varepsilon_a (\tau)|\dot{\varepsilon}_b(\tau) \rangle \ .
\end{eqnarray}
Thus, to derive the lowest order non-adiabatic corrections to our master equations is a matter of repeating our calculations of subsections~\ref{sec:Method1} and \ref{sec:RWA} with $U^{\textrm{ad}}_S(t,t')$ replaced everywhere by the first order term $U^{\textrm{ad}}_S(t,t')V_1(t,t')$, and adding the result to the zeroth order master equations we have already derived. Rather than actually computing these corrections, let us estimate when they are important.

The condition under which the zeroth order adiabatic approximation is accurate is Eq.~\eqref{eq:ad-cond}, which is now replaced by 
\beq
\frac{h}{t_f}\lesssim \Delta^2\ ,
\eeq
i.e., with the $\ll$ replaced by a mere $\lesssim$. However, we would still like to perform the approximation of Eq.~\eqref{eq:U-approx}, in the sense that $U_{S}(t-\tau ,0)\approx e^{i\tau H_{S}(t)}U^{\textrm{ad}}_S(t,t')\left[\ident+ V_1(t,t')\right]$. This still requires 
\beq
\frac{h}{t_f}\ll \frac{1}{\tau_B^2}\ ,
\label{eq:h/t_f}
\eeq 
which is what allows the use of $e^{i\tau H_{S}(t)}$ in this approximation (as shown in Appendix~\ref{app:ShortTime}). 
Taken together, these two conditions are weaker than Eq.~\eqref{eq:strongad}, which we can rewrite as ${\frac{h}{t_f}} \ll \min(\Delta^2,\frac{1}{\tau^2_B})$.

Recall that to ensure the validity of the Markov approximation and $\|\mathcal{L}_{\text{uni}}\| \gg \|\mathcal{L}_{\text{diss}}^{\text{ad}}\|$, we also had to demand the inequalities in Eqs.~\eqref{eq:cond2}, \eqref{eq:Markov-cond}; these can be now summarized as $\max\left(g,\frac{g^2}{\Delta}\right) \ll \frac{1}{\tau_B}$, to be compared with Eq.~\eqref{eq:h/t_f}.


\section{An Illustrative Example: Transverse Field Ising Chain coupled to a Boson Bath}
\label{sec:example}

%

\subsection{The Model} 
\label{sec:Model}

\begin{figure}[ht]
\centering
\includegraphics[width=3in]{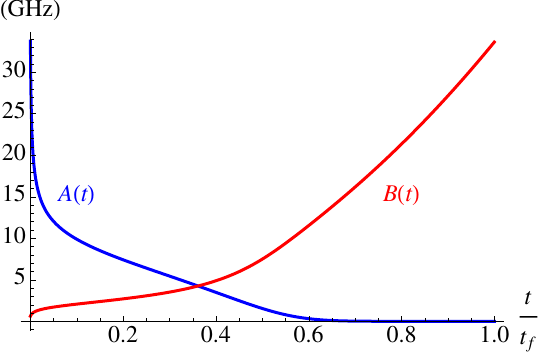}  
\caption{{\protect\small The $A(t)$ and $B(t)$ functions in
Eq.~\eqref{eqt:H_Sa}. $A(0)=33.7$GHz.}}
\label{fig:A_B}
\end{figure}

We proceed to use our master equations to study the evolution of the Ising Hamiltonian with
 transverse field 
\bes
\label{eqt:H_S}
\bea
\label{eqt:H_Sa}
{H_S(t)} &=& A(t) H_S^X + B(t) H_S^Z {\ , } \\
H_S^X &\equiv& \sum_{i=1}^N \sigma_i^x { \ ,}\\
H_S^Z &\equiv&  - \sum_{i=1}^N h_i
\sigma_i^z + \sum_{i,j=1}^N  J_{i j} \sigma_i^z \sigma_j^z 
\ ,
\eea
\ees
where the functions $A(t)$ and $B(t)$ are shown in Fig.~\ref{fig:A_B}, and were chosen {for concreteness} to describe the interpolation between the transverse field and Ising term in the D-Wave Rainier chip \cite{Dwave}.
This is a system which begins with the transverse magnetic field $H_S^X$ turned on while
the Ising term $H_S^Z$ is turned off, and then slowly switches between the two. 

We couple this spin-system to a bath of harmonic
oscillators, with bath and interaction Hamiltonian 
\begin{equation}
\label{eq:SBm}
H_B = \sum_{k=1}^\infty \omega_k b_k^\dagger b_k \ , \quad H_{I} =
\sum_{i=1}^N  \sigma_i^z \otimes B_i \ , \quad B_i = \sum_k g_{k}^i\left(b_k^\dagger + b_k \right) \ ,
\end{equation}
where $b_k^\dagger$ and $b_k$ are, respectively, raising and lowering operators for the $k$th
oscillator with natural frequency $\omega_k$, and $g_{k}^j$ is the corresponding coupling strength to spin $j$. This is the standard pure dephasing spin-boson model \cite{RevModPhys.59.1}, except that our system is time-dependent.  The resulting form for our operator $L$ [Eq.~\eqref{eqt:LindbladL}] is
\begin{equation}
\label{eq:L-ex}
L_{a b, i} = | \varepsilon_a(t) \rangle \langle \varepsilon_a(t) | \sigma^z_{i} | \varepsilon_b(t) \rangle \langle \varepsilon_b(t) | = A_{iab}(t)\ketbra{\varepsilon_a(t)}{\varepsilon_b(t)}\ .
\end{equation}

Recall that our analysis assumed that the bath is in thermal equilibrium at inverse temperature $\beta = 1/(k_BT)$, and hence is described by a thermal Gibbs state $\rho_B = \exp\left( - \beta H_B \right) / \mathcal{Z}$. We show in Appendix~\ref{app:Bath} that this yields
\bes
\bea
\label{eqt:gamma2}
\gamma_{ij} (\omega) &=& \frac{2 \pi J(|\omega|)}{1 - e^{-\beta|
\omega|}} g_{|\omega|}^{i} g_{|\omega|}^{j} \left(
\Theta(\omega)+ e^{-\beta |\omega|} \Theta(-\omega) \right) \\
\label{eq:S_ij}
S_{ i j}(\omega_{ba}(t)) &=& \int_0^\infty d \omega \frac{J(\omega)}{1 -
e^{-\beta \omega}} g_\omega^{i } g_\omega^j \left( \mathcal{P} \left( \frac{1%
}{\omega_{ba}(t) - \omega} \right) + e^{- \beta \omega }\mathcal{P} \left( 
\frac{1}{\omega_{ba}(t) + \omega} \right) \right) \ ,
\eea
\ees
{where only one of the Heaviside functions is non-zero at $\omega =0$.}
{To complete the model specification,} we assume an Ohmic
bath spectral function 
\begin{equation}
J(\omega) = {\eta} \omega
e^{-\omega/\omega_c} \ ,
\label{eq:J-ohmic}
\end{equation}
where $\omega_c$ is a high-frequency cut-off and $\eta$ is a positive constant with dimensions of time squared.

It is often 
stated that the terms associated with the Lamb shift $H_{\textrm{LS}}$ [Eq.~\eqref{eqt:H_LS}], i.e., Eq.~\eqref{eq:S_ij}, can be neglected, since the relative  order of $S$ and $H_S$ is $g^2 \tau_B /\Delta$, and indeed we have assumed $g^2 \tau_B /\Delta \ll 1$ [Eq. \eqref{eq:cond2}]. However, this argument is misleading for two reasons. First, $S$ can be divergent, as is easy to see in the limit $\omega_c=\infty$ for the Ohmic model~\eqref{eq:J-ohmic}, where for $\omega \gg \max_{ba}\omega_{ba}(t)$, the integrand tends to a constant. Second, $S$ should be compared to $\gamma$, as both are of the same order $g^2\tau_B$, and both result in changes to the system relative to its unperturbed state. Indeed, in the interaction picture with respect to $H_S+H_{\textrm{LS}}$ [recall Eq.~\eqref{eqt:RWA}], the overall transition rates between states with quantum numbers $a$ and $b$ will depend on the dressed (i.e., shifted) energy gaps $\omega_{ba}+\omega_{ba}^{\textrm{LS}}$. The importance of this Lamb shift effect was also stressed by de Vega \textit{et al.} \cite{Vega:2010fk}. We analyze the Lamb shift effect in subsection~\ref{subsec:num}.

Finally, we note that although the harmonic oscillators bath with linear coupling to the system provides a $\gamma$ matrix that satisfies the KMS condition,
it is important to note that this model has infrared singularities that destroy the ground state of the total system \cite{Alicki:2004:PDQ:985188.985200}.  The KMS condition assumes a stable ground state and stable thermal states, which our underlying spin-boson model violates.  However, for the purposes of our work, a $\gamma$ matrix that satisfies the KMS condition will suffice without too much concern about how it is derived.

\subsection{{Correlation function analysis}}
\label{subsec:cor-analysis}
\begin{figure}[ht]
\begin{center}
\subfigure[\ $\omega_c = 32 \pi$GHz]{\includegraphics[width=3.0in]{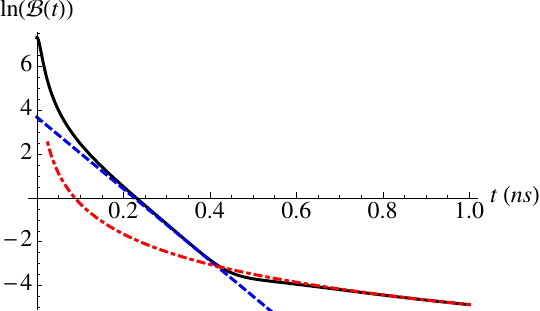}} 
\subfigure[\ $\omega_c = 8\pi$GHz]{\includegraphics[width=3.0in]{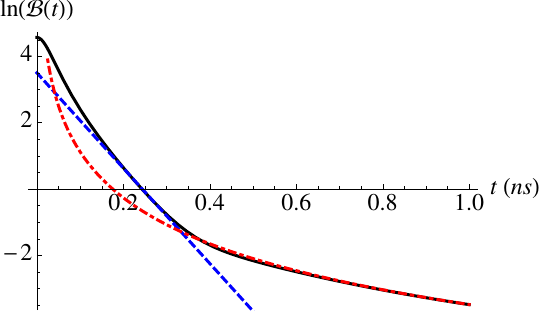}}   
\end{center}
\caption{{\protect\small (a) An example of {the bath correlation function} $\mathcal{B}_{ab}(\tau )$ for $\beta = 2.6^{-1}$ns and $\omega_c = 32\pi$GHz.  The solid black curve is the function $\ln | \mathcal{B}_{ab}(\tau )|$.  The blue dashed curve is the function $\ln 4 \pi^2 \exp (-2 \pi \tau / \beta)$ and the red dot-dashed curve is the function $\ln 2 \beta \tau^{-2}/\omega_c$. (b) Same as in (a), but $\omega_c = 8 \pi$GHz, and the blue dashed curve is the function $\ln d_0 \exp (-(2 \pi / \beta + c_1 \omega_c + c_2 \omega_c^2 \beta )\tau)$ with $c_1 \approx -0.039$,  $c_2 \approx -0.004$, $d_0 \approx 33.13$. }}
   \label{fig:PolyGamma}
\end{figure}

{In light of the subtleties alluded to in subsection~\ref{subsec:KMS} associated with satisfying the KMS condition, we analyze the different timescales determining the behavior of the Ohmic correlation function in this subsection. Removing the $\omega$ dependence from the $g$'s in Eq.~\eqref{eqt:gamma2} and substituting $J(\omega)$ from Eq.~\eqref{eq:J-ohmic}, it is possible to compute the bath correlation function analytically for the resulting}
\beq
\gamma_{ab}(\omega) = \frac{2\pi \omega e^{-|\omega|/\omega_c}}{1-e^{-\b\omega}}g^{a} g^{b}\ ,
\eeq 
by inverse Fourier transform of Eq.~\eqref{eq:gamma_def}. The result is
\beq
\mathcal{B}_{ab}(\tau )= \frac{\eta}{\b^2} g^{a} g^{b} \left( \psi^{(1)} \left(\frac{1}{\b\omega_c} + \frac{i \tau}{\beta} \right) + \psi^{(1)} \left(1 + \frac{1}{\b\omega_c} - \frac{i \tau}{\beta} \right)  \right) \ ,
\label{eq:B-PG}
\eeq
where $\psi^{(m)}$ is the $m$th Polygamma function (see Appendix~\ref{app:PG} for the derivation).  We first {assume 
\beq
\b\omega_c \gg 1\ ,
\label{eq:large-bo}
\eeq
}and consider an expansion in large $\b\omega_c$:
\beq
\mathcal{B}_{ab}(\tau) = \frac{\eta}{\b^2} g^{a} g^{b} \left( - \pi^2 \mathrm{csch}^2\left(\frac{\pi \tau}{\beta} \right) + \sum_{n=1}^{\infty} \frac{\psi^{(n+1)} \left( 1 - \frac{i \tau}{\beta} \right) + \psi^{(n+1)} \left( \frac{i \tau}{\beta} \right)}{n! \left(\beta \omega_c\right)^n}\right) \ ,
\label{eq:67}
\eeq
followed by an expansion in large $\tau/\b$:
\beq
\mathcal{B}_{ab}(\tau) = \frac{\eta}{\b^2} g^{a} g^{b} \left( -4 \pi^2 e^{-\tau/\tau_B} + \frac{1}{(\tau/\tau_M)^2} + O \left( e^{-2\tau/\tau_B} , \tau^{-3} \right)  \right) \ .
\label{eq:Bab-expand}
\eeq
This expansion reveals the two independent time-scales that are relevant for us.  First, there is the time scale $\tau_B$ associated with the exponential decay (corresponding to the true Markovian bath), given by:
\beq
\tau_B \overset{\omega_c \rightarrow\infty}{\longrightarrow } \tau'_B = \frac{\beta}{2 \pi} \ ,
\label{eq:tauB'}
\eeq
then the time scale associated with non-Markovian corrections  (the power law tail):
\beq
\tau_M = \sqrt{ \frac{2 \beta}{\omega_c}} \ .
\eeq
For sufficiently large $\omega_c$, these two time scales capture the two behaviors found in $\mathcal{B}_{ab}(\tau)$, as illustrated in Fig.~\ref{fig:PolyGamma}(a).  

The transition between the exponential decay and the power law tail occurs at a time $\tau_{\mathrm{tr}}$ given by
{${4 \pi^2} e^{-\tau_{\mathrm{tr}} / \tau'_B}
=  (\frac{\tau_M}{\tau_{\mathrm{tr}}})^{2}$,}
{or equivalently by
\beq
 \frac{\exp(\theta)}{\theta^2} = \frac{1}{2}{\b \omega_c}\ , \qquad \theta\equiv 2\pi\frac{\tau_{\mathrm{tr}}}{\b}\ .
 \label{eq:theta}
\eeq
This transcendental equation has a formal solution in terms of the Lambert-$W$ function \cite{Lambert-W}, i.e., the inverse function of $f(W) = W e^W$, as can be seen by changing variables to $y=-\theta/n$, and rewriting $\theta^n e^{-\theta} =a\equiv 2/(\b\omega_c)$ as $ye^y=-a^{1/n}/n$, whose solution is $\theta=-ny = -nW(-\frac{1}{n}a^{1/n})$. However, for our purposes the following observations will suffice. We seek a Markovian-like solution, where $\tau_{\mathrm{tr}}$ is large compared to the thermal timescale set by $\b$, i.e., we are interested in the regime where $\theta \gg 1$. In this case we can neglect $\theta^2$ compared to $e^\theta$, and approximate the solution to Eq.~\eqref{eq:theta} by $\theta \sim \ln(\beta\omega_c/2)$. Thus
\beq
\tau_{\mathrm{tr}} \sim \b\ln(\b\omega_c) \ .
\label{eq:t_tr_ineq}
\eeq
This agrees with the first term of the asymptotic expansion $W(x) = \ln(x)-\ln\ln(x) +\frac{ \ln\ln(x)}{\ln(x)}+\cdots$, which is accurate for $x\gtrsim 3$ \cite{Lambert-W}.
}%

When 
{$\b \omega_c \gg 1$ is not strictly satisfied,} the exponential regime is less pronounced, and $\tau'_B$ is corrected by powers of $\omega_c$.  By dimensional analysis, the corrections must be of the form:
\begin{equation} \label{eqt:tauB}
\tau_B = \tau'_B + \sum_{n=1}^{\infty} \frac{c_n}{\omega_c^n \beta^{n-1}}
\end{equation}
where $c_n$ are constants of order one that must be fitted [see Fig.~\ref{fig:PolyGamma}(b)]. 

{The implications of this cutoff-induced transition for our perturbation theory inequalities are  explored in Appendix~\ref{app:ShortTime}, where we show that a sufficient condition for the theory to hold is
\beq
\frac{1}{\omega_c\ln(\b\omega_c)} < \min\{2\tau_B,\frac{\tau_B h}{\Delta^2 t_f} +\frac{\tau_B^3 h}{t_f}\}\ ,
\label{eq:cutcond}
\eeq
which can be interpreted as saying that the cutoff should be  the largest energy scale. Equation~\eqref{eq:cutcond} joins the list of inequalities given in Section~\ref{sec:time} as an additional special condition that applies in the Ohmic case, along with $\b\omega_c \gg 1$.
}

\subsection{Numerical Results}
\label{subsec:num}

%
For concreteness, we take $g_{\omega}^i =g$,  {$\omega_c = 8 \pi$GHz, and} $T = 20
$mK $\approx 2.6$GHz (in units such that $\hbar = 1$; this is the operating temperature of the D-Wave Rainier chip \cite{Dwave}), {corresponding to $\tau_B = 0.06$ns for the Ohmic model with infinite cutoff, Eq.~\eqref{eq:tauB'}{, and $\tau_B \approx 0.07$ns for $\omega_c = 8 \pi$ using Eq.~\eqref{eqt:tauB}.  For this value of $\omega_c$ the transition between the exponential and power law regimes is still sharp (see Fig~\ref{fig:PolyGamma}(b)) and occurs at approximately $\tau_{\mathrm{tr}} = 0.33$ns. For these values, we satisfy at least one of the cases from Eq.~\eqref{eq:cutcond}, including numerical prefactors: $\frac{1}{\omega_c \ln \left( \beta \omega_c \right)} <  2 \tau_B$.
}
 
We focus on the $N=8$ site ferromagnetic chain with
parameters: 
\begin{equation}
 h_0 = \frac{1}{4} \ , \quad h_{i>0}=0  \ , \quad  J_{i, i+1} = -1 \ , \quad i = 0, \dots, 7 \ ,
\end{equation}
where we pin the first spin in order to break the degeneracy in the ground
state of the classical Ising Hamiltonian. The system is initialized in the Gibbs state: 
\begin{equation}
\rho_S(t = 0) = \frac{e^{- \beta H_S(0)}}{\mathcal{Z}} \ .
\label{eq:rhoS0}
\end{equation}
To help the numerics, we truncate our system to the lowest $n=18$ levels (out of $256$),
rotating the density matrix into the instantaneous energy eigenbasis at each
time step. The error associated with this is small as long as our evolution
does not cause scattering into higher $n$ states, as we have checked. The forward propagation
algorithm used is an implicit second order Runge-Kutte method called TR-BDF2
with an adaptive time step \cite{Coughran:1983,Bank:1985}. %

Figure~\ref{fig:Overlap} presents our results for the evolution of the
system described in the previous subsection. We computed the overlap between the instantaneous ground state of $H_S(t)$ and the instantaneous density matrix predicted by our two master equations~\eqref{eqt:NRWA} (non-RWA) and \eqref{eqt:RWA} (RWA). Although our two master equations
predict 
different numerical values for this overlap, the
qualitative features of the evolution are the same. We observe a
generic feature of four distinct regions of the evolution: a gapped phase
(labeled ``1'' in figure \ref{fig:Overlap}), an excitation phase (labeled
``2''), a relaxation phase (labeled ``3''), and finally a frozen phase
(labeled ``4''). We elaborate on these regions in the following subsection.
Furthermore, we observe that the larger $t_f$ (therefore more adiabatic)
evolution shows a smaller difference between the two master equations for
the final fidelity. 
{The results in Fig.~\ref{fig:Overlap} illustrate the importance of the Lamb shift.  We find that while its effect is small in the RWA, its effect is relatively large in the non-RWA.}
\begin{figure}[ht]
\begin{center}
\subfigure[\ Including Lamb shift]{\includegraphics[width=3.0in]{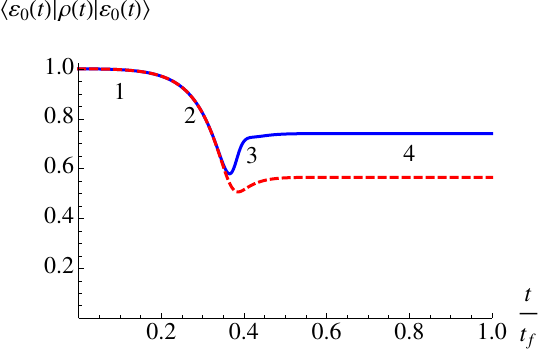}} 
\subfigure[\ Excluding Lamb shift]{\includegraphics[width=3.0in]{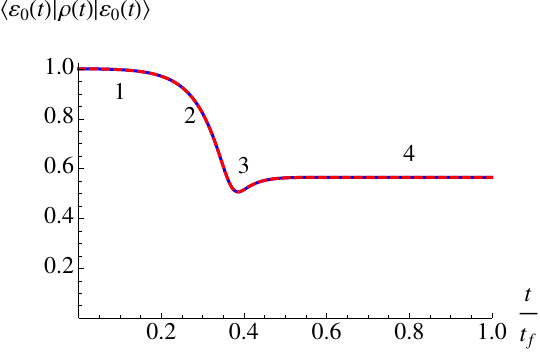}}   
\end{center}
\caption{{\protect\small {Fidelity, or} overlap of the {system} density matrix with the instantaneous ground state along the time evolution for $t_f = 10 \mu s$ and $\eta g^2 / (\hbar^2)= 1.2 \times 10^{-4}/(2 \pi)$. The solid blue curve was calculated
using Eq.~\eqref{eqt:NRWA} (no RWA), while the dashed red curve was calculated 
using Eq.~\eqref{eqt:RWA} (Lindblad form, after the RWA). Four phases are indicated: thermal (1), excitation (2), relaxation (3), and frozen (4). {Panel (a) includes the Lamb shift terms, while (b) excludes them.}}}
\label{fig:Overlap}
\end{figure}

{
In order to study the importance of the relaxation phase, we can study the behavior of the fidelity at $t=t_f$ as we change the coupling strength $g$.  We observe (see Fig.~\ref{fig:coupling}) that there is a rapid drop in fidelity from the closed system result as soon as the coupling to the thermal bath is turned on, but there is a subsequent steady increase in the fidelity as the coupling strength is further increased.  This increase in fidelity is a direct consequence of the 
{higher}
importance of the relaxation phase (made possible by the increasing coupling strength) in restoring the probability of being in the ground state.  {However, we also observe that there is a very pronounced difference between the behavior of the results from the two master equations as the coupling is increased.  In the case of the Lindblad equation, the fidelity saturates, whereas for the NRWA equation, we see an increase in fidelity and a subsequent violation of positivity.  These results bring to light the relative advantages and disadvantages of both master equations.  For the Lindblad equation, positivity of the density matrix is guaranteed, but it clearly is not capturing the physics associated with the increasing importance of the thermal relaxation that the non-RWA equation captures.  However, the non-RWA equation fails to preserve positivity of the density matrix so it is unable to reliably describe the system at higher coupling strength.} {Others have also noted that the RWA and NRWA can lead to physically different conclusions, e.g., in the context of Berry phases in cavity QED \cite{PhysRevLett.108.033601}}.
\begin{figure}[ht]
\begin{center}
\subfigure[\ RWA]{\includegraphics[width=3.0in]{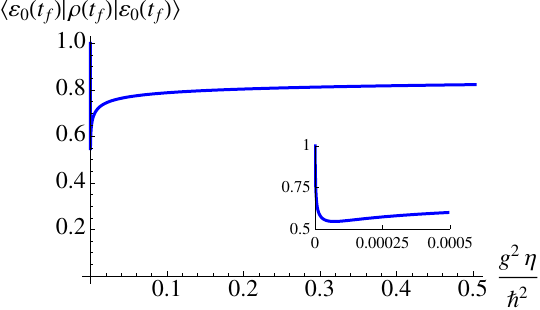}} 
\subfigure[\ non-RWA]{\includegraphics[width=3.0in]{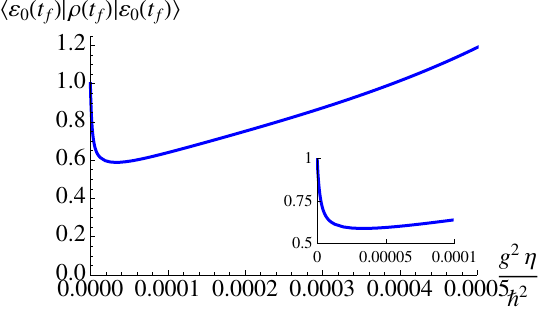}}   
\end{center}
\caption{{\protect\small {Dependence of the} fidelity at $t_f = 10 \mu$s {on} the coupling strength to the thermal bath is varied using our two master equations.  The insets are closeups of the behavior near zero coupling.}}
\label{fig:coupling}
\end{figure}


\subsection{The Four Different Phases}

\subsubsection{Phase 1 -- the gapped phase}
For times sufficiently close to the initial time, the ground state of $H_S(t)$ is the ground state of $H_S^X$, i.e., the state $\ket{0} \equiv \otimes_{j=1}^{N}\ket{-}_j$, where $\ket{\pm}_j = (\ket{\downarrow}_j\pm\ket{\uparrow}_j)/\sqrt{2}$ with energy $\varepsilon_0(0) = -N A(0)$, and where $\ket{\downarrow}_j,\ket{\uparrow}_j$ are the $+1,-1$ eigenstates of $\sigma^z_j$ (computational basis states of the $j$th spin or qubit). The
lowest lying energy states are then the $N$-fold degenerate states with a single flip of one of the spins
in the $x$ direction, i.e., $\ket{i} \equiv \otimes_{j=1}^{i-1}\ket{-}_j\ket{+}_i \otimes_{j=i+1}^{N}\ket{-}_j$, with energy $\varepsilon_1 (0)= -(N-2) A(0)$. Therefore the gap between the ground state and the
first excited states is: 
\begin{equation}
\label{eq:gap}
\Delta(t) = \varepsilon_1(t) - \varepsilon_0(t) \overset{t \ll t_f}{\approx} [-(N-2)
A(t)]-[-N A(t)] = 2 A(t) \ ,
\end{equation}
which is at least twice as large as our $k_B T \approx 2.6$GHz, almost until $A(t_c)=B(t_c) \approx 5$GHz at $t_c \approx 0.35t_f$ (see Fig.~\ref{fig:A_B}). 
Noting that $\sigma_i^z \ket{\pm}_i = \ket{\mp}_i$, we can
write the following relations in terms of the ground and first excited states: 
\begin{equation}
\label{eq:flips}
\omega_{i 0} = \Delta = -\omega_{0i}\ , \quad \sigma_i^z |i \rangle = | 0 \rangle \ .
\end{equation}
{Noting that $\sigma_i^z \ket{j}$ is a doubly excited state $\forall  j\neq i\in\{1,\dots,N\}$, we truncate} the problem to the ground and first excited states only, {so that} there are just three types of values of $\gamma(\omega)$ we need to consider: $\gamma(0),\gamma(\omega_{i0}),\gamma(\omega_{0i})$. Recalling the KMS condition, Eq.~\eqref{eq:KMS}, we have 
\begin{equation}
\label{eq:KMS-ex}
\gamma(\omega_{0i}) = e^{- \beta \omega_{i0}} \gamma(\omega_{i0}) \ ,
\end{equation}
which shows that upward transitions are exponentially suppressed relative to downward transitions, by a factor ranging between $e^{-2A(0)/(k_BT)}\approx e^{-67.4/2.6}\approx 
7 \times 10^{-12} $ and $e^{-2A(0.3t_f)/2.6}\approx e^{-3.8}\approx 0.02$. This explains why for early times (Phase 1) the system hardly deviates from the ground state, which in turn is very close to the thermal state \eqref{eq:rhoS0}. 

\begin{figure}[h]
\begin{center}
\subfigure[\ Ground State]{\includegraphics[width=3.0in]{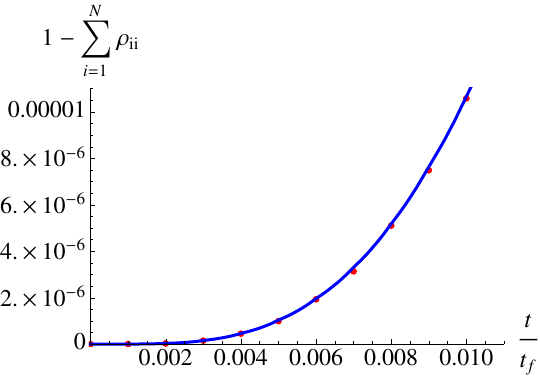}} 
\hspace{0.5cm} \subfigure[\ Excited
states]{\includegraphics[width=3.0in]{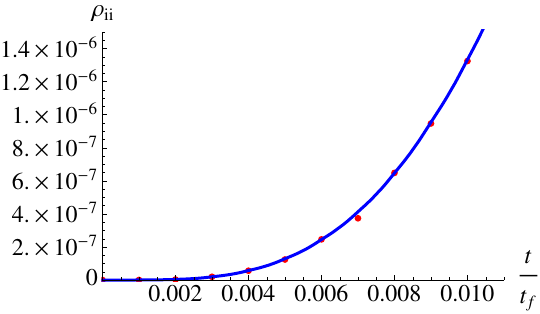}}  
\end{center}
\caption{{\protect\small Early evolution of diagonal elements of the {system} density
matrix in the thermal regime. Red dots are from using the Lindblad equation \eqref{eqt:RWA},
while the solid blue curve is from the solution of Eq.~\eqref{eqt:Thermal_Delta}.}}
\label{fig:Thermal_Phase}
\end{figure}

\begin{figure}[h]
\begin{center}
\subfigure[\; The gap]{\includegraphics[width=3.0in]{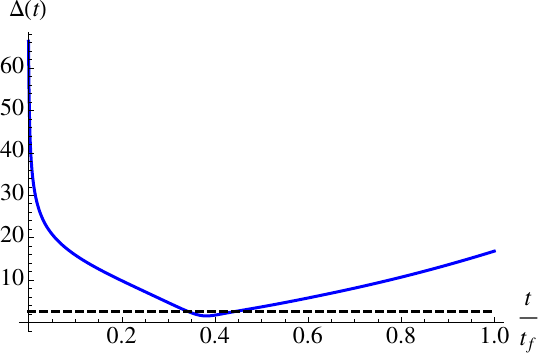}} \hspace{0.5cm} %
\subfigure[\; Suppression
factor]{\includegraphics[width=3.0in]{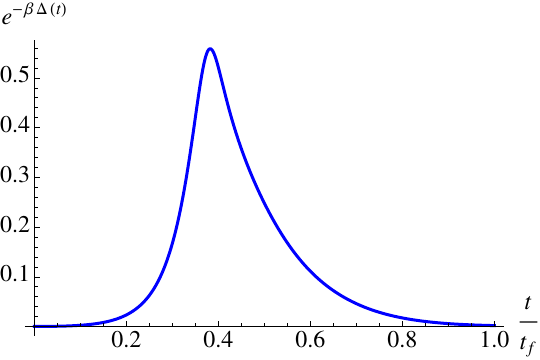}}
\label{fig:KMS_Condition}  
\end{center}
\caption{{\protect\small The behavior of the gap and the exponential
suppression factor along the time evolution. The dashed line is $k_BT \approx 2.6$GHz.}}
\label{fig:Scattering_Phase}
\end{figure}
 
To make this argument more precise, denoting $\rho_{00} = \langle 0 | \rho | 0 \rangle$ and $\rho_{ii} =
\langle i | \rho | i \rangle$, we can write the effective (truncated to the ground and first excited states) Lindblad equation \eqref{eqt:RWA} as the simplified rate equations
\bes
\label{eqt:Thermal_Delta}
\begin{eqnarray}  
\dot{\rho}_{ii} &\approx& \gamma_{i i} (\omega_{i 0}) \left(- \rho_{ii} +
e^{- \beta \omega_{i 0}} \rho_{00} \right) \ , \\
\rho_{00} &\approx&1 - N \rho_{ii} \ ,
\label{eq:rho_00}
\end{eqnarray}
\ees
where we have assumed that the system is initially in the thermal state \eqref{eq:rhoS0} and the gap
is large (relative to $k_B T$).  A derivation of Eq.~\eqref{eqt:Thermal_Delta} can be found in Appendix~\ref{app:Thermal}.
We compare our simulation results
with the results from the above equations in Fig.~\ref{fig:Thermal_Phase}
and find very good agreement for early times.

\subsubsection{Phase 2 -- the excitation phase}\label{subsec:excitation_phase}

When $A(t)$ becomes small enough such that $\beta \Delta(t) \sim O(1)$, then
the KMS condition no longer suppresses excitations from the ground state to
higher excited states (see Fig.~\ref{fig:Scattering_Phase}).  {If we interpret the master equation as a set of rate equations for the matrix elements of $\rho$, we can identify the rate of scattering into the $i$th state from the $j$th state as being the term in the $\dot{\rho}_{ii}$ equation with coefficient $\rho_{jj}$.  Therefore, we find that the rate of scattering from the
ground state to excited states is given by
\begin{equation}
\mathrm{Excitation \ rate} \propto \gamma(\Delta(t)) e^{-\beta \Delta(t)}
\rho_{00} \ ,
\label{eq:excitrate}
\end{equation}
whereas the relaxation rate is given by
\begin{equation}
\mathrm{Relaxation \ rate} \propto \gamma(\Delta(t)) \rho_{ii} \ .
\label{eq:relaxrate}
\end{equation}
Therefore, as we emerge from the gapped phase, we have $\rho_{00} \gg \rho_{ii}$, so
scattering into excited states dominates over relaxation into the
ground state. This explains the loss of fidelity in Phase 2.}

\subsubsection{Phase 3 -- the relaxation phase}

As the gap begins to grow again and the suppression factor shrinks (see Fig.~\ref{fig:Scattering_Phase}), the KMS condition begins to suppress scattering into higher excited states
while allowing relaxation to occur. {In our model} this causes a resurgence in the overlap
with the ground state.  
Therefore, in this phase, the presence of the thermal bath {can actually help to} increase the fidelity, as was also observed in Ref.~\cite{PhysRevLett.100.060503,patane_adiabatic_2009,Vega:2010fk}

The excitation and relaxation phases reveal the two competing processes for a successful adiabatic computation.  If we spend too long in the excitation phase (or if the gap shrinks too fast relative to $t_f$ and the evolution is not adiabatic), the system loses almost all fidelity with the ground state, and the system would have to spend a very long time in the relaxation phase to recover some of that fidelity.  

{However, we stress that fidelity recovery will not be observed if the population becomes distributed over a large number of excited states in the excitation phase. This would happen, e.g., if when the gap closes there is an exponential number of states close to the ground state, such as in the quantum Ising chain with alternating sector interaction defects~\cite{reichardt_quantum_2004}. To see this more explicitly in the context of our analysis,  note from Eq.~\eqref{eq:rho_00} that if there is an exponential number $N$ of $\rho_{ii}$ coupled to $\rho_{00}$ (i.e., $N$ is an exponentially large fraction of the dimension of the system Hilbert space), then
$\rho_{00}$ decreases exponentially. By Eq.~\eqref{eq:excitrate} this means that all $\rho_{ii}$ become
exponentially small, but not zero.
In phase 3, by Eq.~\eqref{eq:relaxrate}, the relaxation is suppressed as long as the gap is not
very large, because the relaxation is proportional to $\rho_{ii}$, which
is exponentially small. This analysis presumes that the system-bath Hamiltonian has non-negligible coupling between the ground and excited state, i.e., that $|\langle 0 | \sigma^z_{\beta} | i \rangle|>0$ in our model [see Eq.~\eqref{eq:J6a}]. This suggests another mechanism that can suppress relaxation: the ground state in phase 1 might have
a large Hamming distance from the ground state in phase 3. Relaxation is then suppressed simply because the coupling is small. 
}

We might expect that there exists an optimum $t_f$ for which the fidelity is maximized by the end of the relaxation phase.  However, for the simple case of the spin-chain we considered, we did not observe such an optimum $t_f$.  The fidelity continues to grow (albeit slowly) for sufficiently high $t_f$. {This is illustrated in Fig.~\ref{fig:LongTime}.}  

\begin{figure}[ht] %
   \centering
   \includegraphics[width=3in]{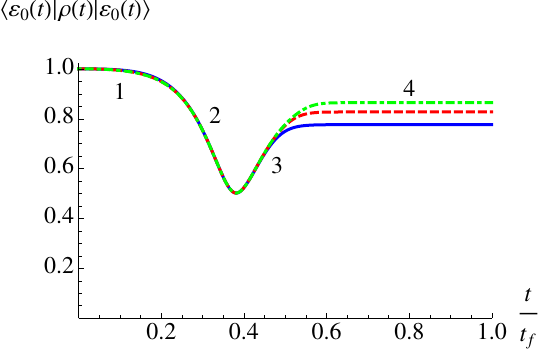} 
   \caption{\protect\small Time evolution {of the system density matrix} using the RWA equation with $\eta g^2 / (\hbar^2)= 0.4/(2 \pi)$ for different $t_f$'s.  {Solid blue curve is for $t_f = 10 \mu$s, dashed red curve is for $t_f = 100 \mu$s}, and dot-dashed green curve is for $t_f = 1$ms. }
   \label{fig:LongTime}
\end{figure}

\subsubsection{Phase 4 -- the frozen phase}
As the gap continues to grow, the relaxation phase ends [notice that the tail in Fig.~\ref{fig:Scattering_Phase}(b) is longer than the actual relaxation phase] and the system's dynamics are frozen {in the ground state}. This is
because $H_S$ becomes almost entirely diagonal in the $\sigma^z$ basis, and so
the off-diagonal components of the $L_{ab,i}$ operators vanish (or become very small), i.e., $A_{i ab}(t) = \langle \varepsilon _{a}(t)|\sigma^z_i|\varepsilon
_{b}(t)\rangle \propto \delta_{ab}$ [Eq.~\eqref{eq:L-ex}]
leaving only the diagonal ones. At this point, {while off-diagonal elements may continue to decay,} the system ground state is no longer affected by the bath;
this is the onset of the frozen phase. 

{
\subsection{Thermal equilibration}
An interesting question is whether the system reaches thermal equilibrium throughout the evolution. To answer this we computed the trace-norm distance [defined in Eq.~\eqref{eq:A1-trace-norm}] between the instantaneous system density matrix and the instantaneous Gibbs state $\rho_{\textrm{Gibbs}}(t) = \exp[-\beta H_S(t)]/\mathcal{Z}$, where $\mathcal{Z}=\textrm{Tr}\left[\exp\left( - \beta H_S(t) \right)\right]$ is the partition function, for the Ising chain discussed above. The result is shown in Fig.~\ref{fig:DistanceGibbs}.  The distance is zero at $t=0$ since the system is initialized in the Gibbs state, and then begins to grow slowly as the system transitions from the gapped phase to the excitation phase.  Though not generic, the distance decreases as the gap shrinks while the excitation phase becomes the relaxation phase, and the system returns to a near Gibbs state where the gap is minimum (at $t/t_f \sim 0.4$).  As the gap opens up again in the transition from the relaxation phase to the frozen phase, the distance begins to grow and continues to grow throughout the frozen phase.  As such, the system is quite far from the Gibbs state at the final time. This feature is significant for adiabatic quantum computation, since it shows the potential for preparation of states which are biased away from thermal equilibrium towards preferential occupation of the ground state.
\begin{figure}[ht] %
   \centering
   \includegraphics[width=3in]{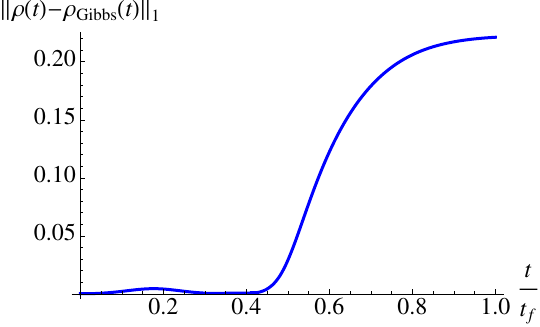} 
   \caption{\protect\small Trace-norm distance between the evolving system density matrix (using the Lindblad equation) and the Gibbs state for an annealing time of $t_f = 10 \mu s$ and $\eta g^2 / (\hbar^2)= 0.4/(2 \pi)$.}
   \label{fig:DistanceGibbs}
\end{figure}
However, we note that on sufficiently long timescales one should expect (from general thermodynamic arguments) terms proportional to $\sigma^x$ and $\sigma^y$ in the system-bath interaction, which we neglected in writing the interaction Hamiltonian $H_I$ in Eq.~\eqref{eq:SBm}, to become important, and to disrupt the frozen phase, allowing the system to fully equilibrate into the Gibbs state. 
}


\section{Conclusions}
\label{sec:conc}

%
Using a bottom-up, first principles approach, we have developed a number of Markovian master equations to describe the adiabatic evolution of a
system with a time-dependent Hamiltonian, coupled to a bath in thermal equilibrium. Our master equations systematically incorporate both time-dependent perturbation theory in the (weak) system-bath coupling $g$, and adiabatic perturbation theory in the inverse of the total evolution time $t_f$. Since we have kept track of the various time- and energy-scales involved in our approximations, higher order corrections (starting at third order in $g$ and second order in $1/t_f$) can be incorporated if desired, a problem we leave for a future publication. Using two of our master
equations, we studied generic features of the adiabatic evolution of a spin chain in the presence of a transverse magnetic field, and coupled to a bosonic heat bath. We identified four phases in this evolution, including a phase where thermal relaxation aids the fidelity of the adiabatic evolution. We hope that this work will prove useful in guiding ongoing experiments on adiabatic quantum information processing, and will serve to inspire the development of increasingly more accurate adiabatic master equations, going beyond the Markovian limit. 

%

\acknowledgments
We are grateful to Robert Alicki for extensive and illuminating discussions regarding the relation between the KMS condition and correlation functions.  We are also grateful to Mohammad Amin for useful discussions and for reading an early version of the manuscript. This research was supported by the ARO MURI grant W911NF-11-1-0268 and by NSF grant numbers PHY-969969 and PHY-803304 (to P.Z. and D.A.L.). 

\appendix

\section{Norms and inequalities}

\label{app:norms} 

We provide a brief summary of norms and inequalities
between them, as pertinent to our work. For more details see, e.g., Refs.~%
\cite%
{Bhatia:1997:SpringerVerlag,Watrous:2004:lecture-notes,Lidar:2008:012308}.
Let $|A|\equiv \sqrt{A^{\dagger }A}$. Unitarily invariant norms are norms
that satisfy, for all unitary $U,V$, and for any operator $A$: $\| UAV\| _{%
\mathrm{ui}}=\| A\| _{\mathrm{ui}}$. Examples of unitarily invariant norms
are the trace norm 
\begin{equation}
\| A\| _{1}\equiv \mathrm{Tr}|A|=\sum_{i}s_{i}(A) \ ,  \label{eq:A1-trace-norm}
\end{equation}%
where $s_{i}(A)$ are the singular values (eigenvalues of $|A|$), and the
supoperator norm, which is the largest eigenvalue of $|A|$: 
\begin{equation}
\| A \| _{\infty }\equiv \sup_{\ket{\psi}:\bra{\psi}\psi\rangle=1} \sqrt{%
\bra{\psi}A^\dagger A\ket{\psi}}=\max_{i}s_{i}(A) \ ,
\end{equation}
Therefore $\| A\ket{\psi}\| \leq \| A \| _{\infty }$ for all normalized
states $\ket{\psi}$, and $\| A\| _{\infty } \leq \| A\| _{1}$.  \newline

All unitarily invariant norms satisfy submultiplicativity: 
\begin{equation}
\| AB\| _{\mathrm{ui}}\leq \| A\| _{\mathrm{ui}}\| B\| _{\mathrm{ui}} \ .
\end{equation}
The norms of interest to us are also multiplicative over tensor products and
obey an ordering: 
\begin{subequations}
\begin{eqnarray}
\| A\otimes B\| _{i} &=&\| A\| _{i}\| B\| _{i}\quad i=1,\infty \ , \\
\| AB\| _{\mathrm{ui}} &\leq &\| A\| _{\infty }\| B\| _{\mathrm{ui}} \ .
\label{eq:ui}
\end{eqnarray}
In particular, $\| AB\|_1 \leq \| A\|_\infty \|B\|_1$.

Another useful fact is that the partial trace is contractive, i.e., 
\end{subequations}
\begin{equation}
\|\text{Tr}_B(X)\|_1, \|\text{Tr}_A(X)\|_1\leq \|X\|_1\ ,
\end{equation}
for any operator $X$ acting on the Hilbert space $\mathcal{H}_A\otimes%
\mathcal{H}_B$. Actually this result extends to other unitarily invariant
norms, with a prefactor depending on the dimension of the traced-out Hilbert
space \cite{Lidar:2008:012308}.

\section{Markov Approximation Bound}

\label{app:MarkovTimeScale} 

%
{We wish to derive an upper bound associated with the error from the
approximation made in Eq.~\eqref{eqt:Markov}, which involves the replacement of $\tilde{\rho}_{S}(t-\tau )$ by $\tilde{\rho}_{S}(t)$ and the extension of the upper integration limit to infinity, i.e., 
\begin{align}
\mc{I}:= \int_{0}^{t}d\tau \left\{A_{\beta }(t-\tau )\tilde{\rho}_{S}(t-\tau )A_{\alpha }(t) + \dots \right\} \mathcal{B}_{\alpha \beta }(\tau ) &\mapsto \int_{0}^{\infty }d\tau \left\{ A_{\beta }(t-\tau )\tilde{
\rho}_{S}(t) A_{\alpha }(t) + \dots \right\} \mathcal{B}_{\alpha \beta }(\tau )\notag \\
& = \mc{I} + \Delta_1 + \Delta_2\ ,
\end{align}
where 
\begin{align}
\Delta_1 &:=\int_{0}^{\infty }d\tau \left\{A_{\beta }(t-\tau )\left[ \tilde{\rho}_{S}(t)-\tilde{\rho}_{S}(t-\tau )\right] A_{\alpha }(t)+ \dots \right\}\mathcal{B}_{\alpha \beta }(\tau ) \notag \\
\Delta_2 &:= \int_{t}^{\infty }d\tau \left\{A_{\beta }(t-\tau )\tilde{\rho}_{S}(t-\tau) A_{\alpha }(t)+ \dots \right\}\mathcal{B}_{\alpha \beta }(\tau ) \ ,
\end{align}
and where the ellipsis denotes the three other summands appearing in Eq.~\eqref{eqt:Markov}. The Markov approximation is more accurate the smaller the error terms $\Delta_1$ and $\Delta_2$.}

{
Consider first the $\Delta_1$ term. We shall show that it is of order $g^2\tau_B^3$. For simplicity we consider only the first of its four summands (the bounds for the other three are identical to that for the first): 
\begin{align}
\left\Vert \int_{0}^{\infty }d\tau A_{\beta }(t-\tau )\left[ \tilde{
\rho}_{S}(t)-\tilde{\rho}_{S}(t-\tau )\right] A_{\alpha }(t)\mathcal{B}%
_{\alpha \beta }(\tau )\right\Vert _{1}
&\leq 
\int_{0}^{\infty }d\tau \|A_{\beta }(t-\tau )\|_\infty \tau\max_{t'\in [t-\tau,t]}\left\Vert \frac{d\tilde{\rho}_{S}(t')}{dt'}\right\Vert_1 \|A_{\alpha }(t)\|_\infty |\mathcal{B}_{\alpha \beta }(\tau )| \notag \\
&=
\int_{0}^{\infty }d\tau\  \tau\max_{t'\in [t-\tau,t]}\left\Vert \frac{d\tilde{\rho}_{S}(t')}{dt'}\right\Vert_1 |\mathcal{B}_{\alpha \beta }(\tau )|\ ,
\label{eq:B5}
\end{align}%
where we used the triangle inequality, $\Vert A_{\alpha}(t)\Vert _{\infty }=\Vert A\Vert =1$ and $\Vert
XY\Vert _{1}\leq \Vert X\Vert _{\infty }\Vert Y\Vert _{1}$ (see Appendix~\ref{app:norms}). We can now use Eq.~\eqref{eqt:eom1} to upper-bound the time derivative:
\begin{align}
\left\Vert\frac{d}{dt}\tilde{\rho}_{S}(t)\right\Vert_1 &\leq {g^{2}}\sum_{\alpha ,\beta
}\int_{0}^{t}d\tau \left( \| A_{\beta }(t-\tau ) \|_\infty \|\tilde{\rho}_{S}(t-\tau
)\|_1 \|A_{\alpha }(t)\|_\infty + \dots \right) |\mathcal{B}_{\alpha \beta }(\tau )|\notag \\
&= 4{g^{2}}\sum_{\alpha ,\beta}\int_{0}^{t}d\tau |\mathcal{B}_{\alpha \beta }(\tau )| \ ,
\end{align}
where the factor of $4$ is due to the same number of summands appearing in Eq.~\eqref{eqt:eom1}.
Substituting this bound back into Eq.~\eqref{eq:B5} we have
\begin{align}
\int_{0}^{\infty }d\tau\  \tau\max_{t'\in [t-\tau,t]}\left\Vert \frac{d\tilde{\rho}_{S}(t')}{dt'}\right\Vert_1 |\mathcal{B}_{\alpha \beta }(\tau )| 
&\leq 
4g^2 \sum_{\alpha \beta} \int_{0}^{\infty }d\tau\  \tau\ |\mathcal{B}_{\alpha \beta }(\tau )| \max_{t'\in [t-\tau,t]} \int_{0}^{t'}d\tau ' |\mathcal{B}_{\alpha \beta }(\tau ')|  \notag \\
&\lesssim
4g^2 \sum_{\alpha \beta} \int_{0}^{\infty }d\tau\  \tau\ |\mathcal{B}_{\alpha \beta }(\tau )|  \tau_B  \leq 4g^2 \tau_B^3 \sum_{\alpha \beta} 1
\ ,
\label{eq:B4}
\end{align}
where we used $\int_{0}^{t'}d\tau ' |\mathcal{B}_{\alpha \beta }(\tau ')|  \leq \int_{0}^{\infty}d\tau ' |\mathcal{B}_{\alpha \beta }(\tau ')| $ and Eq.~\eqref{eq:corr-decay}.
}

This is to be compared to the term we use after the Markov
approximation: 
\begin{equation}
\left\Vert \int_{0}^{\infty }d\tau A_{\beta }(t-\tau )_{\beta }\tilde{\rho}%
_{S}(t)A_{\alpha }(t)\mathcal{B}_{\alpha \beta }(\tau )\right\Vert _{1}\leq
\int_{0}^{\infty }d\tau |{\mathcal{B}}_{\alpha \beta }(\tau )|\sim \tau _{B}\ .
\label{eq:B6}
\end{equation}%
Comparing Eqs.~\eqref{eq:B4} and \eqref{eq:B6}, we see that the relative
error is {$O[(g\tau _{B})^2]$}.

{
Next consider the $\Delta_2$ term. We have
\begin{align}
\left\Vert\int_{t}^{\infty }d\tau A_{\beta }(t-\tau )\tilde{\rho}_{S}(t-\tau) A_{\alpha }(t)\mathcal{B}_{\alpha \beta }(\tau) \right\Vert_1 
&\leq 
\int_{t}^{\infty }d\tau \|A_{\beta }(t-\tau )\|_\infty \|\tilde{\rho}_{S}(t-\tau)\|_1 \|A_{\alpha }(t)\|_\infty |\mathcal{B}_{\alpha \beta }(\tau)| \notag \\
&= \int_{t}^{\infty }d\tau\ |\mathcal{B}_{\alpha \beta }(\tau)| \ .
\end{align}
The last integral converges provided $|\mathcal{B}_{\alpha \beta }(\tau)| \sim 1/\tau^x$ with $x>1$. This is typically the case. E.g., for the Ohmic spin-boson model discussed in subsection~\ref{subsec:cor-analysis} we show that for $t>\tau_{\mathrm{tr}} \sim \b\ln(\b\omega_c)$, the bath correlation function $|\mathcal{B}_{\alpha \beta }(\tau)| \sim \frac{\eta }{\b\omega_c} \frac{1}{\tau^2}$. In this case, then, we can set $t = \tau_{\mathrm{tr}}$ and bound
\beq
\|\Delta_2\|_1 \lesssim \frac{\eta}{\b\omega_c} \int_{\tau_{\mathrm{tr}}}^\infty  d\tau\ \frac{1}{\tau^2} \sim \frac{\eta}{\b^2 \omega_c \ln(\b\omega_c)}\ ,
\label{eq:B8}
\eeq
which tends to zero as the cutoff tends to infinity at a fixed finite temperature (even if the lower integration limit is replaced by a constant), as expected in the weak coupling limit assumed in our work. Note that the infinite temperature limit, where Eq.~\eqref{eq:B8} diverges, is incompatible with weak coupling, and requires the so-called singular coupling limit  \cite{Breuer:2002,PhysRevA.73.052311}.}

\section{Properties of the spectral-density matrix $\Gamma_{\alpha\beta}(\omega)$}

\label{app:Gamma} 
Introducing the Fourier transform pair 
\begin{equation}
\mathcal{B}_{\alpha \beta }(\tau ,0)=\int_{-\infty }^{\infty }\frac{d\omega 
}{2\pi }e^{-i\omega \tau }\gamma _{\alpha \beta }(\omega )\ ,\quad \gamma
_{\alpha \beta }(\omega )=\int_{-\infty }^{\infty }d\tau e^{i\omega \tau }%
\mathcal{B}_{\alpha \beta }(\tau ,0)\ ,
\end{equation}%
and using the property that 
\begin{equation}
\int_{0}^{\infty }d\tau e^{i(\omega -\omega ^{\prime })\tau }=\pi \delta
(\omega -\omega ^{\prime })+i\mathcal{P}\left( \frac{1}{\omega -\omega
^{\prime }}\right) \ ,
\end{equation}%
where $\mathcal{P}$ denotes the Cauchy principal value,\footnote{%
By definition, $\mathcal{P}\left( \frac{1}{x}\right) [f]=\lim_{\epsilon
\rightarrow 0}\int_{\mathbb{R}\backslash \lbrack -\epsilon ,\epsilon ]}\frac{%
f(x)}{x}dx$, where $f$ belongs to the set of smooth functions with compact
support on the real line $\mathbb{R}$.} we have, using Eq.~%
\eqref{eqt:SpectralDensity}, 
\begin{equation}
\Gamma _{\alpha \beta }(\omega )=\int_{0}^{\infty }d\tau \ e^{i\omega \tau
}\int_{-\infty }^{\infty }\frac{d\omega ^{\prime }}{2\pi }e^{-i\omega
^{\prime }\tau }\gamma _{\alpha \beta }(\omega ^{\prime })=\int_{-\infty
}^{\infty }\frac{d\omega ^{\prime }}{2\pi }\gamma _{\alpha \beta }(\omega
^{\prime })\int_{0}^{\infty }d\tau \ e^{i(\omega -\omega ^{\prime })\tau }=%
\frac{1}{2}\gamma _{\alpha \beta }(\omega )+iS_{\alpha \beta }(\omega )\ ,
\end{equation}%
in agreement with Eq.~\eqref{eqt:Gamma}, where 
\begin{equation}
S_{\alpha \beta }(\omega )=\int_{-\infty }^{\infty }\frac{d\omega ^{\prime }%
}{2\pi }\gamma _{\alpha \beta }(\omega ^{\prime })\mathcal{P}\left( \frac{1}{%
\omega -\omega ^{\prime }}\right) \ ,
\end{equation}%
in agreement with Eq.~\eqref{eq:gS}. 

Note that: 
\begin{equation}
\mathcal{B}_{\alpha \beta }^{\ast }(\tau ,0)=\langle B_{\alpha }(\tau
)B_{\beta }(0)\rangle ^{\ast }=\mathrm{Tr}\left[ B_{\alpha }(\tau )B_{\beta
}(0)\rho _{B}\right] ^{\ast }=\mathrm{Tr}\left[ B_{\beta }(0)B_{\alpha
}(\tau )\rho _{B}\right] =\mathcal{B}_{\beta \alpha }(0,\tau )\ .
\end{equation}%
When $\rho _{B}$ commutes with $H_{B}$ (which we have assumed), we further
have%
\begin{equation}
\mathcal{B}_{\beta \alpha }(0,\tau )=\mathrm{Tr}\left[ B_{\beta
}(0)U_{B}^{\dag }(\tau )B_{\alpha }(0)U_{B}(\tau )\rho _{B}\right] =\mathrm{%
Tr}\left[ \rho _{B}U_{B}(\tau )B_{\beta }(0)U_{B}^{\dag }(\tau )B_{\alpha
}(0)\right] =\mathcal{B}_{\beta \alpha }(-\tau ,0)\ .
\end{equation}
Thus the spectral-density matrix satisfies: 
\begin{equation}
\Gamma _{\alpha \beta }^{\ast }(\omega )=\int_{0}^{\infty }d\tau \
e^{-i\omega \tau }\mathcal{B}_{\alpha \beta }^{\ast }(\tau
,0)=\int_{0}^{\infty }d\tau \ e^{-i\omega \tau }\mathcal{B}_{\beta \alpha
}(-\tau ,0)=\int_{-\infty }^{0}d\tau \ e^{i\omega \tau }\mathcal{B}_{\beta
\alpha }(\tau ,0)=\frac{1}{2}\gamma _{\beta \alpha }(\omega )-iS_{\beta
\alpha }(\omega )\ .
\end{equation}


\section{Proof of the KMS condition}
\label{app:KMS}

The proof of the time-domain version of the KMS condition, Eq.~\eqref{eq:KMSt}, is the following calculation: 
\bes
\bea
\langle B_a(\tau)B_b(0)\rangle &=& \textrm{Tr}[\rho_B U^\dag_B(\tau,0) B_a U_B(\tau,0) B_b] = \frac{1}{\mathcal{Z}}\textrm{Tr}[B_b e^{-(\beta-i\tau)H_B} B_a e^{-i\tau H_B}] \\
&=& \frac{1}{\mathcal{Z}}\textrm{Tr}[B_b e^{i(\tau+i\beta)H_B} B_a e^{-i(\tau+i\beta) H_B}e^{-\beta H_B}]
= \textrm{Tr}[\rho_B B_b U^\dag_B(\tau+i\beta,0) B_a U_B(\tau+i\beta,0) ] \\
&=& \langle B_b(0)B_a(\tau+i\beta)\rangle\ .
\eea 
\ees
Note that using the same technique it also follows that \beq
\langle B_a(\tau)B_b(0)\rangle = \langle B_{b}(-\tau - i \beta) B_{a}(0) \rangle\ .
\eeq 

To prove that this implies the frequency domain condition, Eq.~\eqref{eq:KMS}, we compute the Fourier transform:
\bea
\gamma_{ab} (\omega) = \int_{-\infty}^{\infty} d \tau e^{i \omega \tau} \langle B_a(\tau)B_b(0)\rangle 
= \int_{-\infty}^{\infty} d \tau e^{i \omega \tau} \langle B_{b}(-\tau - i \beta) B_{a}(0) \rangle
\label{eqt:integrand1}
\eea
%
%
\begin{figure}[h]
   \includegraphics[width=2in]{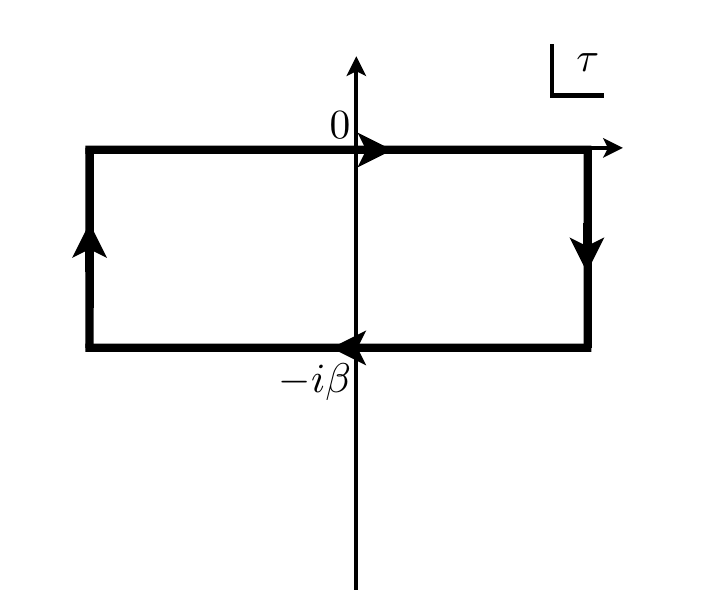} 
   \caption{Contour used in proof of the KMS condition.}    
   \label{fig:Contour}
\end{figure}
To perform this integral we replace it with a contour integral in the complex plane, 
$\oint_C d \tau e^{i \omega \tau} \langle B_{b}(-\tau - i \beta) B_{a}(0) \rangle $,
with the contour $C$ as shown in Fig.~\ref{fig:Contour}. This contour integral vanishes by the Cauchy-Goursat theorem \cite{complex:book} since the closed contour encloses no poles (the correlation function $\langle B_{b}(\tau) B_{a}(0)\rangle$ is analytic in the open strip $(0, -i \beta)$ and is continuous at the boundary of the strip \cite{KMS}), so that
\begin{equation}
\oint_{C} \left( \dots \right) = 0 = \int_{\uparrow} \left( \dots \right) + \int_{\mathrm{\downarrow}}  \left( \dots \right)  + \int_{\rightarrow}  \left( \dots \right)  +  \int_{\leftarrow}  \left( \dots \right) 
\end{equation}
where $\left( \dots \right)$ is the integrand of Eq.~\eqref{eqt:integrand1}, and the integral $ \int_{\rightarrow}$ is the same as in Eq.~\eqref{eqt:integrand1}.  After making the variable transformation $\tau = -x - i \beta$, where $x$ is real, we have
\begin{equation}
\int_{\leftarrow}  \left( \dots \right)  = -  e^{\beta \omega}  \int_{-\infty}^{\infty} e^{-i \omega x}  \langle B_{b}(x) B_{a}(0) \rangle
\end{equation}
Assuming that $\langle B_a(\pm\infty-i\b)B_b(0)\rangle = 0$ (i.e., the correlation function vanishes at infinite time), we further have $\int_{\uparrow} \left( \dots \right) = \int_{\mathrm{\downarrow}}  \left( \dots \right) =0$, and hence we find the result:
\begin{equation}
\int_{-\infty}^{\infty} d \tau e^{i \omega \tau} \langle B_{b}(-\tau - i \beta) B_{a}(0) \rangle =  e^{\beta \omega}  \int_{-\infty}^{\infty} e^{-i \omega \tau}  \langle B_{b}(\tau) B_{a} (0) \rangle  = e^{\beta \omega} \gamma_{ba}(-\omega)\ ,
\end{equation}
which, together with Eq.~\eqref{eqt:integrand1}, proves Eq.~\eqref{eq:KMS}.

\section{Non-Adiabatic Corrections} 
For completeness we provide a brief review of pertinent aspects of adiabatic perturbation theory, and the derivation of the adiabatic condition relating the total evolution time to the ground state gap. See, e.g., Ref.~\cite{Teufel:book} for additional details.

\subsection{Adiabatic perturbation theory}
\label{app:AdC}
%
%
To consider adiabatic corrections, we recall that  $U_S$ satisfies
\begin{equation}
\label{eq:C1}
i \partial_t U_S(t,t') = H_S(t) U_S(t,t') \ ,
\end{equation}
where 
\beq
{H}_S(t) =  \sum_a \varepsilon_a(t) \ketbra{\varepsilon_a(t)}{\varepsilon_a(t)} \ .
\eeq
Define the ``adiabatic intertwiner" $W(t,t')$:
\begin{equation}
\label{eq:C3}
W(t,t') \equiv \sum_a \ketbra{\varepsilon_a(t)}{\varepsilon_a(t')} = T_+ \exp[-i\int_{t'}^t d\tau \ K(\tau)]\ ,
\end{equation}
where the ``intertwiner Hamiltonian" is
\bes
\bea
K(t) &\equiv& i [\partial_t W(t,t')]W^\dagger(t,t') = i\sum_{a} \ketbra{\dot{\varepsilon}_a(t)}{\varepsilon_a(t)}\\
&=& i[\dot{P}_0(t),P_0(t)]\ ,
\label{eq:Avron}
\eea
\ees
and where $P_0(t) \equiv \ketbra{\varepsilon_0(t)}{\varepsilon_0(t)}$ is the projection onto the ground state of $H_S(t)$. The result \eqref{eq:Avron} is by no means obvious and is proven in subsection~\ref{sec:Avron}.

To extract the geometric phase we define
\bes
\bea
H_G(t) &\equiv& \sum_a \phi_a(t) \ketbra{\varepsilon_a(t)}{\varepsilon_a(t)} \ , \\
\phi_a(t) &=& i\bracket{\varepsilon_a(t)}{\dot{\varepsilon}_a(t)}\ , \\
H'_S(t) &\equiv& H_S(t)-H_G(t) \ .
\label{eq:H'S}
\eea
\ees

Now define $V$ via the ``adiabatic interaction picture" transformation:
\begin{equation}
V(t,t') = W^\dagger(t,t')U_S(t,t')  \ ,
\end{equation}
along with
\bes
\bea
\tilde{H}_S(t,t') &=& W^\dagger(t,t') H_S(t) W(t,t') = \sum_a \varepsilon_a(t) \ketbra{\varepsilon_a(t')}{\varepsilon_a(t')} \ , \\
\tilde{H}_G(t,t') &=& W^\dagger(t,t') H_G(t) W(t,t') = \sum_a \phi_a(t) \ketbra{\varepsilon_a(t')}{\varepsilon_a(t')} \ , \\
\tilde{H}'_S(t,t') &\equiv& \tilde{H}_S(t,t')-\tilde{H}_G(t,t')  \ , \\
\tilde{K}(t,t') &=& W^\dagger(t,t') K(t) W(t,t') = iW^\dagger(t,t') \partial_t W(t,t')\ .
\eea
\ees
Note that the time dependence of $\tilde{H}_S$ and $\tilde{H}'_S(t)$ is entirely in the energy eigenvalue and not in the eigenstates.  
Then $V$ obeys the Schr\"odinger equation:
\begin{equation} 
\label{eqt:dtUtilde}
i \partial_t V(t,t') = \tilde{H}_S^{\textrm{ad}}(t,t') V(t,t') \ .
\end{equation}
where
\beq
\tilde{H}_S^{\textrm{ad}}(t,t') \equiv \tilde{H}_S(t,t') - \tilde{K}(t,t') = W^\dagger(t,t') [H_S(t)-K(t)] W(t,t')\ .
\eeq
When the evolution is nearly adiabatic $\tilde{H}_S^{\textrm{ad}}(t,t')$ is a perturbation, so that we  consider a solution of Eq.~\eqref{eqt:dtUtilde} for $V$ of the form:
\begin{equation} \label{eqt:AdiabaticExpansion}
V(t,t') = V_0(t,t') \left( \mathds{1} + V_1(t,t') + \dots \right) \ ,
\end{equation}
with the zeroth order solution associated with the purely adiabatic evolution, including the geometric phase:
\bes
\bea
V_0(t,t') &\equiv& T_+ \exp \left[-i \int_{t'}^t d \tau \tilde{H}'_S(\tau,t')\right] \\ 
\label{eq:USad}
U^{\textrm{ad}}_S(t,t') &\equiv& W(t,t') V_0(t,t') = \sum_a | \varepsilon_a(t) \rangle \langle \varepsilon_a (t')| e^{-i \int_{t'}^t d \tau \left[\varepsilon_a(\tau)-\phi_a(\tau)\right] } \ .
\eea
\ees
Differentiating with respect to $t$, we have
\bes
\bea
\dot{U}^{\textrm{ad}}_S(t,t') &=& \dot{W}(t,t')V_0(t,t')+W(t,t')\dot{V}_0(t,t') \\
&=& -iK(t)W(t,t')V_0(t,t')-iW(t,t')\tilde{H}'_S(t,t')V_0(t,t')\\
 &=& -i\left[H_S^{\textrm{ad}}(t)-H_G(t)\right]U^{\textrm{ad}}_S(t,t') \ ,
\eea
\ees
where
\bea
\label{eq:Had}
H_S^{\textrm{ad}}(t) &\equiv& K(t)+H_S(t) \ .
\eea

Plugging the Eq.~\eqref{eqt:AdiabaticExpansion} expansion into Eq.~\eqref{eqt:dtUtilde}, we obtain, to first order in $1/t_f$:\footnote{To see that this is an expansion in powers of $1/t_f$, transform to the dimensionless time variable $s=t/t_f$.}
\bes
\bea
i \partial_t V_1 (t,t')  &=& - V^\dagger_0(t,t') \left[\tilde{K}(t,t')-\tilde{H}_G(t,t')\right] V_0(t,t') \\
&=& 
-i U^{\textrm{ad}\dag}_S(t,t') \partial_t W(t,t') V_0(t,t') +U^{\textrm{ad}\dag}_S(t,t')H_G(t)U^{\textrm{ad}}_S(t,t') \ .
\eea
\ees
Note that $U^{\textrm{ad}\dag}_S(t,t')H_G(t)U^{\textrm{ad}}_S(t,t') = \sum_a\phi_a(t) \ketbra{\varepsilon_a(t')}{\varepsilon_a(t')}$. Therefore, integrating, and using 
\beq
\mu_{ba}(t,t') = \int_{t'}^t  d\tau \, \omega_{ba}(\tau)\ , \qquad \omega_{ba}(\tau) = [\varepsilon_b(\tau)-\phi_b(t)]  - [\varepsilon_a(\tau) -\phi_a(t)]\ ,
\eeq
we can write the solution as:
\begin{eqnarray}
V_1(t,t') &=& - \int_{t'}^t d \tau\ \left[U^{\textrm{ad}\dag}_S(\tau,t') \partial_\tau W(\tau , t') V_0(\tau, t') -\sum_a\phi_a(\tau) \ketbra{\varepsilon_a(t')}{\varepsilon_a(t')} \right] \nonumber \\
&=&- \sum_{a \neq b} \int_{t'}^t d \tau e^{- i \mu_{ba}(\tau,t')} |\varepsilon_a(t') \rangle \langle \varepsilon_b(t') | \langle \varepsilon_a (\tau)|\dot{\varepsilon}_b(\tau) \rangle \ .
\end{eqnarray}
Therefore, the system evolution operator can be written as:
\begin{equation}
\label{eq:ad1st}
U_S(t,t') = U^{\textrm{ad}}_S(t,t') + Q(t,t') U^{\textrm{ad}}_S(t,t') + O\left(t_f^{-2}\right) \ ,
\end{equation}
where the correction term can be made appropriately dimensionless in a more careful second order analysis, and where 
\begin{equation} 
\label{eqt:Q}
Q(t,t') \equiv U^{\textrm{ad}}_S(t,t') V_1(t,t') U^{\textrm{ad}\dagger}_S(t,t') = \sum_{a \neq b} e^{-i \mu_{ab}(t,t')} \left( - \int_{t'}^t d \tau e^{i \mu_{ab}(\tau,t')} \langle \varepsilon_a (\tau)|  \dot{\varepsilon}_b(\tau) \rangle \right) | \varepsilon_a (t) \rangle \langle \varepsilon_b(t)|  \ .
\end{equation}

\subsection{Adiabatic timescale analysis}
\label{app:adiabatic-time}
Equation~\eqref{eq:ad1st} shows that the first order correction to the purely adiabatic evolution is given by $Q$. Thus the adiabatic timescale is found by ensuring that the matrix elements of $Q$ are all small. Let us show that a sufficient condition for this is 
$\frac{h}{\Delta^2 t_f} \ll 1$ [Eq.~\eqref{eq:ad-cond}]. More rigorous analyses replace the $\ll$ condition with an inequality relating the same parameters to the fidelity between the ground state of $H_S(t_f)$ and the solution of the Schr\"odinger at $t_f$, e.g., \cite{Jansen:07,Boixo:10,lidar:102106,Wiebe:12}.

Starting from Eq.~\eqref{eqt:Q} and taking matrix elements we have
\begin{equation} 
Q_{ab}(t,t') =   e^{-i \mu_{ab}(t,t')} \left( - \int_{t'}^t d \tau e^{i \mu_{ab}(\tau,t')} \langle \varepsilon_a (\tau)| \partial_\tau | \varepsilon_b(\tau) \rangle \right)  \ .
\end{equation}
Changing variables to dimensionless time $s=t/t_f$, $\mu_{ab}(t,t')$ becomes $ t_f \int_{t'/t_f}^{s}  ds' \, \omega_{ab}(s') = t_f \tilde{\mu}_{ab}(s,t'/t_f)$, and 
\begin{equation} \label{eqt:integrand}
\int_{t'}^t d \tau e^{i \mu_{ab}(\tau,t')} \langle \varepsilon_a (\tau)| \partial_\tau | \varepsilon_b(\tau) \rangle = \int_{t'/t_f}^{s} ds' \ e^{i t_f \tilde{\mu}_{ab}(s',t'/t_f)}  \langle \varepsilon_a (s')| \partial_{s'} | \varepsilon_b(s') \rangle \ ,
\end{equation}
where $\tilde{\mu}$ now involves a dimensionless time integration.  Using the fact that $e^{ i t_f \tilde{\mu}_{ab} (s',t'/t_f) } = \frac{-i}{t_f \omega_{ab}(s')} \frac{d}{d s'} e^{ i t_f \tilde{\mu}_{ab} (s',t'/t_f)}$,
we can integrate Eq.~\eqref{eqt:integrand} by parts as
\begin{eqnarray}
\label{eq:C17}
\int_{t'/t_f}^{s} ds' \ e^{i t_f \tilde{\mu}_{ab}(s',t'/T)}  \langle \varepsilon_a (s')| \partial_{s'} | \varepsilon_b(s') \rangle &=& \frac{i}{t_f \omega_{ab}(s')} e^{i t_f \tilde{\mu}_{ab}(s',t'/t_f)} \langle \varepsilon_a(s')|\partial_{s'} | \varepsilon_b(s') \rangle \Big|_{t'/t_f}^{s} \nonumber \\
&& - \frac{i}{t_f} \int_{t'/t_f}^{s} ds' \ e^{i t_f \tilde{\mu}_{ab}(s',t'/t_f)} \frac{d}{d s'} \frac{\langle \varepsilon_a(s')|\partial_{s'} | \varepsilon_b(s') \rangle}{\omega_{ab}(s')} \ .
\end{eqnarray}
Now note that for non-degenerate energy eigenstates, differentiation with respect to $t$ of $H_S(t)\ket{\varepsilon_a(t)} = \varepsilon_a(t)\ket{\varepsilon_a(t)}$, and substitution of $s=t/t_f$ in Eq.~\eqref{eq:h}, directly yields the relation:\footnote{Differentiating (where $\dot{x}\equiv \partial_t x$), we have $\dot{H}_S\ket{\varepsilon_a(t)}+H_S(t)\ket{\dot{\varepsilon}_a(t)} = \dot{\varepsilon}_a(t)\ket{\varepsilon_a(t)}+\varepsilon_a(t)\ket{\dot{\varepsilon}_a(t)}$. Taking matrix elements gives $\bra{\varepsilon_b(t)}\dot{H}_S\ket{\varepsilon_a(t)}+\varepsilon_b(t)\bracket{\varepsilon_b(t)}{\dot{\varepsilon}_a(t)}=\dot{\varepsilon}_a(t)\delta_{ab}+\varepsilon_a(t)\bracket{\varepsilon_b(t)}{\dot{\varepsilon}_a(t)}$, which yields Eq.~\eqref{eq:<dot>} provided $\ket{\varepsilon_b(t)}$ is an eigenstate non-degenerate with $\ket{\varepsilon_a(t)}$.}
\begin{equation}
 \bracket{\varepsilon_a(t)}{\dot{\varepsilon}_b(t)}  = \frac{ \langle \varepsilon_a (t) | \dot{H}_S(t) | \varepsilon_b (t) \rangle }{\omega_{ba} (t)} \sim \frac{h}{\Delta t_f} \ .
 \label{eq:<dot>}
\end{equation}

Substitution into Eq.~\eqref{eq:C17} yields, with the help of Eq.~\eqref{eq:<dot>},
\beq
|Q_{ab}(t,t')| \leq \frac{|\langle \varepsilon_a (s) | \partial_s H_S(s) | \varepsilon_b (s) \rangle |}{\omega_{ab}^2(s) t_f}  \Big|_{t'/t_f}^{t/t_f} \nonumber + \left| \int_{t'/t_f}^{t/t_f} ds' \ \frac{e^{i t_f \tilde{\mu}_{ab}(s',t'/t_f)}}{ t_f}\frac{d}{d s'} \frac{\langle \varepsilon_a (s') | \partial_{s'} H_S(s') | \varepsilon_b (s') \rangle} {\omega_{ab}^2(s')}\right| \ .
\eeq
Continued integration by parts will yield higher powers of 
the dimensionless quantity $ \frac{\langle \varepsilon_a (s) | \partial_\tau H_S(s) | \varepsilon_b (s) \rangle}{\omega_{ab}^2(s) t_f}$. Thus a sufficient condition for the smallness of $|Q_{ab}(t,t')|$ for all $a,b$ is that 
\beq
\frac{\max_{s;a,b} |\langle \varepsilon_a (s) | \partial_s H_S(s) | \varepsilon_b (s) \rangle |}{\min_{s;a,b}\omega_{ab}^2(s) t_f} \ll 1\ ,
\eeq
namely the condition in Eq.~\eqref{eq:ad-cond}, where we assumed that the minimal Bohr frequency is the ground state gap, $\omega_{1,0}$.
Of course, this argument is by no means rigorous, and in fact it has been the subject of considerable discussion, e.g., \cite{PhysRevLett.93.160408,Sarandy:04,Tong:07,Amin:09,PhysRevLett.106.138902,0253-6102-57-3-03}. For rigorous analyses see, e.g., Refs.~\cite{Teufel:book,Jansen:07,PhysRevA.78.052508,lidar:102106,Boixo:10,Wiebe:12}. For our purposes Eq.~\eqref{eq:ad-cond} suffices.

\subsection{Intertwiner Hamiltonian}
\label{sec:Avron}
Here we provide an explicit proof of Eq.~\eqref{eq:Avron}, a well-known result due to Avron et al. \cite{Avron:87}. The original proof lacks some detail, so our purpose here is to fill in the gaps, for completeness of our presentation (see also Ref.~\cite{PhysRevA.77.042319} for details of the proof). Define the ground state and orthogonal projectors $P_0(t) \equiv \ketbra{\varepsilon_0(t)}{\varepsilon_0(t)}$ and $Q_0(t) \equiv \ident-P_0(t)$. Recalling Eq.~\eqref{eq:ad1st}, we have
\begin{equation}
\label{}
P_0(t) U_S(t,t') P_0(t') = P_0(t) U^{\textrm{ad}}_S(t,t')P_0(t') + P_0(t) Q(t,t') U^{\textrm{ad}}_S(t,t')P_0(t')  + O\left(t_f^{-2}\right)  \ .
\end{equation}
Using Eq.~\eqref{eq:USad} we find $U^{\textrm{ad}}_S(t,t')P_0(t') = {P_0(t)} e^{-i \int_{t'}^t d \tau \varepsilon_0(\tau) }$, so that up to a phase $P_0(t) Q(t,t') U^{\textrm{ad}}_S(t,t')P_0(t') = P_0(t) Q(t,t') P_0(t')$. However, $Q(t,t')$ [Eq.~\eqref{eqt:Q}] contains only off-diagonal terms [since we subtracted the Berry connection in Eq.~\eqref{eq:H'S}], so that $P_0(t) Q(t,t') P_0(t')=0$. 
Thus, to first order in $1/{t_f}$ the evolution generated by $H_S(t)$ keeps the ground state decoupled from the excited states, i.e., 
\bes
\bea
P_0(t) U_S(t,t') P_0(t') &=& U^{\textrm{ad}}_S(t,t') P_0(t') + O\left(t_f^{-2}\right) {\ ,} \\
Q_0(t) U_S(t,t') Q_0(t') &=& U^{\textrm{ad}}_S(t,t') Q_0(t') + O\left(t_f^{-2}\right) {\ .}
\eea
\ees
Letting $\tau = t-t' >0$ and expanding around $t$, we then have, using Eqs.~\eqref{eq:C1} and \eqref{eq:C3},
\bes
\bea
P_0(t) [\ident + i\tau H_S(t)][P_0(t) -\tau \dot{P}_0(t)] &=& [\ident + i\tau  H_S^{\textrm{ad}}(t)][P_0(t) -\tau \dot{P}_0(t)] + O\left(t_f^{-2}\right) {\ ,} \\
Q_0(t) [\ident + i\tau H_S(t)][Q_0(t) -\tau \dot{Q}_0(t)] &=& [\ident + i\tau  H_S^{\textrm{ad}}(t)][Q_0(t) -\tau \dot{Q}_0(t)] + O\left(t_f^{-2}\right) \ .
\eea
\ees
Using the projector properties $P_0 = P_0^2 \Longrightarrow \dot{P}_0 P_0 + P_0\dot{P}_0 = \dot{P}_0$, and $[H_S(t), P_0(t)] =0$  (similarly for $Q_0$), this simplifies to
\bes
\bea
-\tau P_0(t)\dot{P}_0(t) + i\tau H_S(t) P_0(t) &=& -\tau \dot{P}_0(t) + i\tau H_S^{\textrm{ad}}(t) P_0(t)+O\left(t_f^{-2}\right)  \ , \\ 
-\tau Q_0(t)\dot{Q}_0(t) + i\tau H_S(t) Q_0(t) &=& -\tau \dot{Q}_0(t) + i\tau H_S^{\textrm{ad}}(t) Q_0(t)+O\left(t_f^{-2}\right) \ ,
\eea
\ees
and, after dropping the $O\left(t_f^{-2}\right)$ corrections, to
\bes
\begin{align}
\label{eq:23a}
&i[H_S^{\textrm{ad}}(t)-H_S(t)]P_0(t) + \dot{P}_0(t) P_0(t) = 0 \ , \\
&i[H_S^{\textrm{ad}}(t)-H_S(t)]Q_0(t) + \dot{Q}_0(t) Q_0(t) = 0 \ .
\label{eq:23b}
\end{align}
\ees
Inserting $Q_0(t) = \ident - P_0(t)$ and $\dot{Q}_0(t) = -\dot{P}_0(t)$ into Eq.~\eqref{eq:23b}, and subtracting it from Eq.~\eqref{eq:23a}, we find, using $\dot{P}_0 P_0 + P_0\dot{P}_0 = \dot{P}_0$ once more, the desired result:
\beq
H_S^{\textrm{ad}}(t) = H_S(t) + i[\dot{P}_0(t), P_0(t)]\ ,
\eeq
which, together with Eq.~\eqref{eq:Had}, proves Eq.~\eqref{eq:Avron}.

%
\section{Short Time Bound} 
\label{app:ShortTime}

We wish to bound the error associated with neglecting $\Theta$ in Eq.~\eqref{eq:U-approx}, i.e., we wish to bound
\beq
\|\Theta (t,\tau )\|_\infty = \|U_{S}(t-\tau ,0)-e^{i\tau H_{S}(t)}U_{S}^{\text{ad}}(t,0)\|_\infty = \|U_{S}^{\text{ad}\dagger}(t,0)e^{-i\tau H_{S}(t)}U_S^\dagger(t,t-\tau)U_S(t,0)-\ident\|_\infty  \ .
\eeq
Using that the operator $\hat{U}(\tau) = e^{-i\tau H_{S}(t)}U_S^\dagger(t,t-\tau) $ satisfies: 
\begin{equation}
\frac{d}{d \tau} \hat{U}(\tau)= \left( H_S(t) - e^{-i H_S(t) \tau} H_S(t-\tau) e^{i H_S(t) \tau} \right) \hat{U}(\tau) \ ,
\end{equation}
we can write a formal solution for $\hat{U}$ as:
\begin{equation}
\hat{U}(\tau) = \ident - i \int_0^{\tau} dt' \left[ H_S(t) - e^{-i H_S(t) t'} H_S(t-t') e^{i H_S(t) t'} \right] \hat{U}(t') \ . 
\end{equation}
Therefore we can bound:
\bes
\bea
\|\Theta (t,\tau )\|_\infty  &=& \left\Vert U_{S}^{\text{ad}\dagger}(t,0)U_S(t,0) - i \int_0^\tau d t' U_{S}^{\text{ad}}(t,0)\left[ H_S(t)  - e^{-i H_S(t) t'} H_S(t-t') e^{i H_S(t) t'} \right] \hat{U}(t')U_S(t,0)-\ident\right\Vert_\infty \\
&\leq& {\min\{2,\|Q(t,0)\|_\infty + \int_0^\tau d t' \|\left[ H_S(t)  - e^{-i H_S(t) t'} H_S(t-t') e^{i H_S(t) t'} \right] \|_\infty + O(t_f^{-2})\}}
\eea
\ees
%
where we used Eq.~\eqref{eq:ad1st} {and the fact that supoperator norm between two unitaries is always upper bounded by $2$} in the second line, and the standard adiabatic estimate to bound $\|Q(t,0)\|_\infty$ (recall subsection~\ref{app:adiabatic-time}). While $h$ of Eq.~\eqref{eq:h} is expressed in terms of a matrix element, a more careful analysis (e.g., Ref.~\cite{lidar:102106}) would replace this with an operator norm. Thus we shall make the plausible assumption that $h \sim t_f\max_{t'\in[t-\tau,t]}\|\partial_{t'} H_S(t') \|_\infty$, and, dropping the subdominant $O(t_f^{-2})$, we can write
\beq 
\|\Theta (t,\tau )\|_\infty \leq {\min\{2,\frac{h}{\Delta^2 t_f} +\frac{\tau^2 \tilde{h}}{2t_f}\}}\ .
\label{eq:E5}
\eeq
where $\tilde{h}=\max_{t'\in[t-\tau,t]} \frac{t_f}{t'} \|\left[ H_S(t)  - e^{-i H_S(t) t'} H_S(t-t') e^{i H_S(t) t'} \right] \|_\infty$.
We can now bound the error term in Eq.~\eqref{eq:33b}. Let $X(t,\tau)\equiv e^{i\tau H_{S}(t)}U_{S}^{\text{ad}}(t,0)$, so that $U_{S}(t-\tau ,0)=X(t,\tau)+\Theta(t,\tau)$. We can then write
\bes
\begin{align}
&\int_{0}^{\infty }d\tau A_{\beta }(t-\tau )\tilde{\rho}_{S}(t)A_{\alpha }(t)%
\mathcal{B}_{\alpha \beta }(\tau ) =\int_{0}^{\infty }d\tau X^\dagger(t,\tau)A_{\beta }X(t,\tau)\tilde{\rho}_{S}(t)A_{\alpha }(t)\mathcal{B}_{\alpha \beta }(\tau )
\label{eq:E6a}
\\
&\quad+
\int_{0}^{\infty }d\tau X^\dagger(t,\tau)A_{\beta }\Theta(t,\tau)\tilde{\rho}_{S}(t)A_{\alpha }(t)\mathcal{B}_{\alpha \beta }(\tau )+
\int_{0}^{\infty }d\tau X\Theta^\dagger(t,\tau)A_{\beta }X(t,\tau)\tilde{\rho}_{S}(t)A_{\alpha }(t)\mathcal{B}_{\alpha \beta }(\tau )
\label{eq:E6b}\\
&\quad +\int_{0}^{\infty }d\tau \Theta^\dagger(t,\tau)A_{\beta }\Theta(t,\tau)\tilde{\rho}_{S}(t)A_{\alpha }(t)\mathcal{B}_{\alpha \beta }(\tau )\ .
\label{eq:E6c}
\end{align}
\ees
The first term on the RHS of \eqref{eq:E6a} is the approximation we have used in Eq.~\eqref{eq:33b}. The terms in \eqref{eq:E6b} and \eqref{eq:E6c} can be bounded as follows, using Eq.~\eqref{eq:E5}, the unitarity of $X$, the fact that $\|\tilde{\rho}_S\|_\infty \leq 1$, and recalling that $\|A_\alpha\|_\infty=1$. {First we assume that Eq.~\eqref{eq:corr-decay} applies. Then:}
\bes
\label{eq:E7}
\begin{align}
&\left\Vert\int_{0}^{\infty }d\tau X^\dagger(t,\tau)A_{\beta }\Theta(t,\tau)\tilde{\rho}_{S}(t)A_{\alpha }(t)\mathcal{B}_{\alpha \beta }(\tau )\right\Vert_\infty ,  \left\Vert\int_{0}^{\infty }d\tau X\Theta^\dagger(t,\tau)A_{\beta }X(t,\tau)\tilde{\rho}_{S}(t)A_{\alpha }(t)\mathcal{B}_{\alpha \beta }(\tau )\right\Vert_\infty \notag \\
&\quad \leq \int_{0}^{\infty }d\tau \|\Theta(t,\tau)\|_\infty |\mathcal{B}_{\alpha \beta }(\tau )| {\lesssim \min\{2\tau_B,\frac{\tau_B h}{\Delta^2 t_f} +\frac{\tau_B^3 \tilde{h}}{2 t_f}\}} \ , \\
&\left\Vert\int_{0}^{\infty }d\tau \Theta^\dagger(t,\tau)A_{\beta }\Theta(t,\tau)\tilde{\rho}_{S}(t)A_{\alpha }(t)\mathcal{B}_{\alpha \beta }(\tau )\right\Vert_\infty \leq \int_{0}^{\infty }d\tau \|\Theta(t,\tau)\|^2_\infty |\mathcal{B}_{\alpha \beta }(\tau )| \lesssim {\min\{4\tau_B,2 \frac{\tau_B h}{\Delta^2 t_f} +2 \frac{\tau_B^3 \tilde{h}}{2t_f}\}}\ ,
\label{eq:E7b}
\end{align}
\ees
{where in the last inequality we used the fact that if $x \leq 2$ then $[\min(2,x)]^2 = x^2 \leq 2x$, and if $x \geq 2$ then again $[\min(2,x)]^2 = 2\min(2,x) \leq 2x$, with $x=\frac{h}{\Delta^2 t_f} +\frac{\tau^2 h}{t_f}$, in order to avoid having to extend Eq.~\eqref{eq:corr-decay} to higher values of $n$. In all, then, the approximation error in Eq.~\eqref{eq:33b} is $O[\min\{\tau_B,\frac{\tau_B h}{\Delta^2 t_f} + \frac{\tau_B^3 \tilde{h}}{t_f}\}]$.}

{Next we recall from the discussion in subsection~\ref{subsec:KMS} that Eq.~\eqref{eq:corr-decay} must, in the case of a Markovian bath with a finite cutoff, be replaced by the weaker condition~\eqref{eq:corr-decay2}, reflecting fast decay up to $\tau_{\mathrm{tr}}$, followed by power-law decay. In this case the terms in \eqref{eq:E6b} can instead be bounded as follows:
\begin{align}
&\left\Vert\int_{0}^{\infty }d\tau X^\dagger(t,\tau)A_{\beta }\Theta(t,\tau)\tilde{\rho}_{S}(t)A_{\alpha }(t)\mathcal{B}_{\alpha \beta }(\tau )\right\Vert_\infty ,  \left\Vert\int_{0}^{\infty }d\tau X\Theta^\dagger(t,\tau)A_{\beta }X(t,\tau)\tilde{\rho}_{S}(t)A_{\alpha }(t)\mathcal{B}_{\alpha \beta }(\tau )\right\Vert_\infty \notag \\
&\quad \leq \int_{0}^{\tau_{\mathrm{tr}}}d\tau \|\Theta(t,\tau)\|_\infty |\mathcal{B}_{\alpha \beta }(\tau )|+ \int_{\tau_{\mathrm{tr}}}^{\infty}d\tau \ \|\Theta(t,\tau)\|_\infty |\mathcal{B}_{\alpha \beta }(\tau )| \notag \\
&\quad \lesssim \min\{2\tau_B,\frac{\tau_B h}{\Delta^2 t_f} +\frac{\tau_B^3 \tilde{h}}{2 t_f}\} + 2\frac{\tau_M^2}{\tau_{\mathrm{tr}}}\ ,
\label{eq:E8}
\end{align}
in place of \eqref{eq:E7b}. A similar modification can be computed for the term in \eqref{eq:E6c}. To compute the order of $\frac{\tau_M^2}{\tau_{\mathrm{tr}}}$, we recall that $\tau_{\mathrm{tr}} \sim \b\ln(\b\omega_c) \gg \b$ [Eq.~\eqref{eq:t_tr_ineq}], and that $\tau_M =\sqrt{2 \beta/\omega_c}$. 
Thus $\frac{\tau_M^2}{\tau_{\mathrm{tr}}} \sim [\omega_c\ln(\b\omega_c)]^{-1}$. It follows that we can safely ignore the $\frac{\tau_M^2}{\tau_{\mathrm{tr}}}$ term in Eq.~\eqref{eq:E8} provided Eq.~\eqref{eq:cutcond} is satisfied. The analysis of the term in \eqref{eq:E6c} does not change this conclusion.
}


\section{Derivation of the Schr\"{o}dinger picture adiabatic master equation in Lindblad form}
\label{app:RWA}

Starting from Eq.~\eqref{eq:46} and performing a transformation back to the Schr\"{o}dinger picture, along with a double-sided adiabatic approximation, yields
\bes
\begin{align}
&U_S(t,0)\int_{0}^{t}d\tau A_{\beta }(t-\tau )\tilde{\rho}_{S}(t)A_{\alpha }(t)
\mathcal{B}_{\alpha \beta }(\tau )U^\dag_S(t,0)+\textrm{h.c.}\approx  \\
&\quad   \sum_{\omega}A_{\beta, \omega}(t)A_{\alpha,\omega}(t)\Pi_{\omega}(t)
{\rho}_{S}(t)\Pi_{\omega}(t) \Gamma_{\alpha\beta}(\omega)  +\textrm{h.c.}=  \sum_{\omega}L_{\omega,\beta }(t)
{\rho}_{S}(t)L^\dag_{\omega,\alpha }(t) \Gamma_{\alpha\beta}(\omega) +\textrm{h.c.}\ .
\label{eq:45c}
\end{align}
\ees

Similarly, the other terms yield 
\beq
-U_S(t,0)\int_{0}^{t}d\tau A_{\alpha }(t)A_{\beta }(t-\tau )\tilde{\rho}_{S}(t)
\mathcal{B}_{\alpha \beta }(\tau )U^\dag_S(t,0)\approx  
 -\sum_{\omega}L^\dag_{\omega,\alpha }(t)L_{\omega,\beta }(t)
{\rho}_{S}(t)\Gamma_{\alpha\beta}(\omega) -\textrm{h.c.}   \ .
\label{eq:46b}
\eeq
Using Eqs.~\eqref{eqt:Gamma} and \eqref{eq:gS} for the spectral-density matrix and its complex conjugation, we are able to combine the Hermitian conjugate terms, starting from the terms in Eq.~\eqref{eq:45c}:
\bes
\begin{align}
&\sum_{\alpha \beta} \Gamma_{\alpha \beta} (\omega) L_{\omega, \beta} (t)  \rho_S (t) L^\dagger_{\omega, \alpha}(t)  + \mathrm{h.c.} \\
&\qquad =\sum_{\alpha \beta} \left[\Gamma_{\alpha \beta} (\omega) L_{\omega, \beta} (t)  \rho_S (t) L^\dagger_{\omega, \alpha}(t) + \Gamma^{\ast}_{\alpha \beta} (\omega) L_{\omega, \alpha} (t)  \rho_S (t) L^\dagger_{\omega, \beta}(t) \right]  \\
&\qquad =\sum_{\alpha \beta} \gamma_{\alpha \beta}(\omega) L_{\omega, \beta} (t)  \rho_S (t) L^\dagger_{\omega, \alpha}(t) \ ,
\end{align}
\ees
where in the last equality we used the freedom to interchange $\alpha$ and $\beta$ under the summation sign. Likewise, the terms in Eq.~\eqref{eq:46b} yield
\bes
\begin{align}
&\sum_{\alpha \beta} \Gamma_{\alpha \beta} (\omega) L^\dagger_{\omega, \alpha} (t)  L_{\omega, \beta}(t) \rho_S (t)   + \mathrm{h.c.} \\
&\qquad = \sum_{\alpha \beta} \left[\Gamma_{\alpha \beta} (\omega) L^\dagger_{\omega, \alpha} (t)  L_{\omega, \beta}(t) \rho_S (t) + \Gamma^\ast_{\alpha \beta} (\omega) \rho_S (t)  L^\dagger_{\omega, \beta}(t) L_{\omega, \alpha} (t) \right]  \\
&\qquad = \sum_{\alpha \beta} \gamma_{\alpha \beta}(\omega) \frac{1}{2} \left\{ L^\dagger_{\omega, \alpha} (t)  L_{\omega, \beta}(t), \rho_S(t) \right\}  + i S_{\alpha \beta}(\omega) \left[L^\dagger_{\omega, \alpha} (t)  L_{\omega, \beta}(t), \rho_S(t) \right]\ ,
\end{align}
\ees
where $\{,\}$ denotes the anticommutator. Combining these results with a similar calculation for the $aa$ and $bb$ terms in Eqs.~\eqref{eq:45c} and \eqref{eq:46b}, finally results in Eqs.~\eqref{eqt:RWA} and \eqref{eqt:H_LS}.


\section{Calculations for the Spin-Boson Model}

\label{app:Bath} 
We recall the following basic facts about the bosonic operators appearing in Eq.~\eqref{eq:SBm}, where $\{n\}\equiv\{n_k\}_k$ is the set of occupation numbers of all modes:
\bes
\begin{eqnarray}
&\left[ b_k, b_{k^{\prime }}^\dagger \right] = \delta_{k k^{\prime }} \ ,
\quad b_k |\{n\} \rangle = \sqrt{n_k} | \{n_k-1, n\} \rangle \ , \quad
b_k^\dagger |\{n\} \rangle = \sqrt{n_k+1} | \{n_k+1, n\} \rangle \ ,&  
\\
& H_B |\{n\} \rangle = E_{\{n\}} |\{n\} \rangle = \left( \sum_k n_k \omega_k
\right) |\{n\} \rangle \ .&
\end{eqnarray}
\ees

%
Recalling that the bath operators $B_i = \sum_k g_{k}^i\left(b_k^\dagger + b_k \right)$, we proceed to calculate $\langle B_{i}(t) B_{j}(0) \rangle = \langle
B_{i}^\dagger(t)B_{j}(0) \rangle $ assuming that the bath is in a thermal Gibbs state $
\rho_B = \exp\left( - \beta H_B \right) / \mathcal{Z}$, where $\mathcal{Z}=\textrm{Tr}\left(\exp\left( - \beta H_B \right)\right)$ is the partition function. We begin by writing: 
\begin{eqnarray}
\langle B_{i}(t) B_{j}(0) \rangle &=& \mathrm{Tr}_B \left( U_B^{\dagger}
(t,0) B_{i}^{\dagger} U_B(t,0) B_j \rho_b \right)  \notag \\
&=& \sum_{ \{m\}, \{n \} , \{p \}} \langle \{ m \} | U_B^{\dagger} (t,0)
B_{i}^{\dagger} | \{n \} \rangle \langle \{ n \} | U_B (t,0) B_{j} | \{p \}
\rangle \langle \{ p \} | \rho_b | \{m \} \rangle \ .
\end{eqnarray}
The time evolution operator acting on the eigenstates simply produces a
phase, so we focus on the operator $B_j$'s role. 
\begin{equation}
\langle \{ n \} | B_{j} | \{p \} \rangle = \sum_k \left( g_k^{j} \sqrt{p_k+1}
\delta_{\{n\}, \{p_k+1,p\}} + g_k^{ j } \sqrt{p_k} \delta_{\{n\},
\{p_k-1,p\}} \right) \ .
\end{equation}
Plugging this result in, we find: 
\begin{eqnarray}
\langle B_{i}(t) B_{j}(0) \rangle &=& \frac{1}{\mathcal{Z}}\sum_{\{m \}, \{ n \} , \{p \} }
\sum_{k,k^{\prime }} e^{i \left(E_{\{m\}} - E_{\{n\}} \right)t} {e^{- j
E_{\{m\}}}}  \notag \\
&&\times \left( g_k^{i} g_{k^{\prime }}^j \sqrt{n_k+1}\sqrt{p_{k^{\prime }}+1%
} \delta_{\{m \}, \{n,n_k+1 \}} \delta_{\{n \},\{p,p_{k^{\prime }}+1 \}}
\delta_{\{p \} \{m \}} \right.  \notag \\
&& + \left. g_k^{i } g_{k^{\prime }}^j \sqrt{n_k}\sqrt{p_{k^{\prime }}+1}
\delta_{\{m \}, \{n,n_k-1 \}} \delta_{\{n \},\{p,p_{k^{\prime }}+1 \}}
\delta_{\{p \} \{m \}} \right.  \notag \\
&& + \left. g_k^{i } g_{k^{\prime }}^{j } \sqrt{n_k+1}\sqrt{p_{k^{\prime }}}
\delta_{\{m \}, \{n,n_k+1 \}} \delta_{\{n \},\{p,p_{k^{\prime }}-1 \}}
\delta_{\{p \} \{m \}} \right.  \notag \\
&& + \left. g_k^{i } g_{k^{\prime }}^{j } \sqrt{n_k}\sqrt{p_{k^{\prime }}}
\delta_{\{m \}, \{n,n_k-1 \}} \delta_{\{n \},\{p,p_{k^{\prime }}-1 \}}
\delta_{\{p \} \{m \}} \right) \ .
\end{eqnarray}
The only terms that are non--zero are the middle two, giving us: 
\begin{equation}
\langle B_{i}(t) B_{j}(0) \rangle = \sum_{\{m \}} \sum_{k} \frac{g_k^{i
} g_k^{j}}{\mathcal{Z}} \left( m_k +1 \right) \left(e^{- \beta
E_{\{m\}}}e^{i \left(E_{\{m\}} - E_{\{m_k+1\}} \right)t} + e^{- \beta
E_{\{m_k+1\}}}e^{i \left(E_{\{m_k+1\}} - E_{\{m\}} \right)t} \right)  \ .
\end{equation}
Using that: 
\begin{equation}
E_{\{m\}} - E_{\{m_k+1\}} = - \omega_k \ , \quad \mathcal{Z} =
\prod_{k=1}^{\infty} \sum_{m_k = 0}^{\infty} e^{- \beta m_k \omega_k} =
\prod_{k=1}^{\infty} \frac{1}{1- e^{- \beta \omega_k}} \ ,
\end{equation}
we can write: 
\begin{equation}
\sum_{\{m \}} \left(m_k + 1\right)e^{- \beta E_{\{m\}}} = 1 - \frac{1}{\beta}
\frac{\partial_{\omega_k}\mathcal{\ Z}}{\mathcal{Z}} = \frac{1}{1- e^{-\beta
\omega_k}} \ .
\end{equation}
We can simplify our expression to 
\begin{equation}
\langle B_{i}(t) B_{j}(0) \rangle = \sum_k \frac{1}{1- e^{- \beta
\omega_k }} \left( e^{-i \omega_k t} g_k^{i } g_k^j + e^{ i \omega_k t-
\beta \omega_k} g_k^{i} g_k^{j } \right) \ .
\end{equation}
We can replace the sum with a 
integral: 
\begin{equation}
\sum_k = \sum_{\omega^{\prime }} \int_0^\infty d\omega f(\omega)
\delta(\omega - \omega^{\prime }) = \int_0^\infty d \omega J(\omega) \ ,
\end{equation}
where $f(\omega)$ is a measure of the number of oscillators at frequency $%
\omega$ and $J(\omega)$ is the bath spectral function. Our final result is
then: 
\begin{equation}
\langle B_{i}(t) B_{j}(0) \rangle = \int_0^\infty d \omega \frac{%
J(\omega)}{1 - e^{-\beta \omega}} \left( e^{-i \omega t} g_\omega^{i }
g_\omega^j + e^{ i \omega t- \beta \omega} g_\omega^{i} g_\omega^{j }
\right) \ .
\end{equation}
where we have assumed that oscillators with the same $\omega_k$ value
interact with the $i^{\textrm{th}}$ spin with the same interaction strength, i.e. 
\begin{equation}
g_k^i = g_{k^{\prime }}^i \ \mathrm{if} \ \omega_k = \omega_{k^{\prime }} \ .
\end{equation}
Plugging our result into Eq.~\eqref{eq:gS}, we find: 
\bes
\bea
\gamma_{i j} (\omega_{ba}(t)) &=& \frac{2 \pi J(|\omega_{ba}(t)|)}{1 -
e^{-\beta| \omega_{ba}(t)|}} \left( g_{|\omega_{ba}(t)|}^{i }
g_{|\omega_{ba}(t)|}^{j} \Theta(\omega_{ba}(t))+ e^{-\beta |\omega_{ba}(t)|}
g_{|\omega_{ba}(t) |}^i g_{|\omega_{ba}(t)|}^{j } \Theta(-\omega_{ba}(t))
\right) \\
S_{ i j}(\omega_{ba}(t)) &=& \int_0^\infty d \omega \frac{J(\omega)}{1
- e^{-\beta \omega}} \left( g_\omega^{i } g_\omega^j \mathcal{P} \left( 
\frac{1}{\omega_{ba}(t) - \omega} \right) + g_\omega^i g_\omega^{j } e^{-
\beta \omega }\mathcal{P} \left( \frac{1}{\omega_{ba}(t) + \omega} \right)
\right) \ ,
\eea
\ees
where $\Theta$ denotes the Heaviside step function.
%

\section{Derivation of the Ohmic Bath Correlation Function}
\label{app:PG}

Here we derive the bath correlation function for the Ohmic oscillator bath with a finite frequency cutoff, Eq.~\eqref{eq:B-PG}.
We start from the simplified expression for the spectral-density,
\beq
\gamma(\omega) = 2\pi {\eta} g^2 \frac{\omega
e^{-|\omega|/\omega_c}}{1-e^{-\b\omega}}\ ,
\eeq 
and compute the bath correlation function by inverse Fourier transform of Eq.~\eqref{eq:gamma_def}:
\beq
\mathcal{B}(\tau ) =  \frac{1}{2\pi}\int_{-\infty}^\infty d \omega\ e^{-i \omega \tau} \gamma(\omega) 
= \frac{{\eta} g^2}{\b^2} \left(\int_{-\infty}^0 dx\ e^{-i x \tau/\b}\frac{x e^{x/(\b \omega_c)}}{1-e^{-x}}
+  \int_{0}^{\infty} dx\ e^{-i x \tau/\b}\frac{x e^{-x/(\b \omega_c)}}{1-e^{-x}}\right)
\ ,
\label{eq:I2}
\eeq
where we changed variables to $x=\b \omega$.

The Polygamma function is defined as $\psi^{(m)}(z) \equiv\frac{d^{m+1}}{dz^{m+1}} \ln\Gamma(z)$, where $\Gamma(z)  = \int_0^\infty  e^{-x} x^{z-1} dx$ is the Gamma function.
The Polygamma function may be represented for $\Re(z)>0$ and $m>0$ as
$\psi^{(m)}(z)= (-1)^{m+1}\int_0^\infty
\frac{x^m e^{-xz}} {1-e^{-x}} dx$, so that in particular, for $m=1$, we have
\beq
\psi^{(1)}(z)= \int_0^\infty
\frac{x e^{-xz}} {1-e^{-x}} dx \ .
\eeq
We can rewrite Eq.~\eqref{eq:I2}
\bes
\bea
\mathcal{B}(\tau ) &=& \frac{{\eta} g^2}{\b^2} \left(\int_0^{\infty} dx\ \frac{x e^{-x[-i\tau/\b+1+1/(\b\omega_c)]}}{1-e^{-x}}
+  \int_0^{\infty} dx\ \frac{x e^{-x[i\tau/\b+1/(\b\omega_c)]}}{1-e^{-x}}\right)\\
&=& \frac{{\eta} g^2}{\b^2} \left( \psi^{(1)}(-i\tau/\b+1+1/(\b\omega_c)) + \psi^{(1)}(i\tau/\b+1/(\b\omega_c)) \right) \ ,
\eea
\ees
which is the desired result.
%

\section{Derivation of the Effective Rate Equations in the Thermal Phase} \label{app:Thermal}
%
Here we derive the rate equation \eqref{eqt:Thermal_Delta}.

\subsection{Single qubit}
For illustration, let us consider a single qubit such that the Hamiltonian is given by
\begin{equation}
H_S(t) = A(t) \sigma^x - B(t) h_z \sigma^z \ .
\end{equation}
The gap at any given time in the evolution is
\begin{equation}
\Delta(t) = 2 \sqrt{ A(t)^2 + B(t)^2 h_z^2 } \ .
\end{equation}
Since there are only two states, there are only three $\gamma(\omega)$ terms to calculate (assuming $g_\omega = g$):
\begin{equation}
\gamma(0) = \frac{2 \pi g^2}{\beta} \ , \quad \gamma(+ \Delta(t) ) = 2 \pi g^2 \frac{\Delta(t)}{1 - e^{- \beta \Delta(t)}} \ , \quad \gamma(- \Delta(t) ) = 2 \pi g^2 \frac{\Delta(t) e^{-\beta \Delta( t)}}{1 - e^{- \beta \Delta(t)}} \ ,
\end{equation}
where we have taken the limit $\omega_c \rightarrow\infty$ for simplicity.
Let us consider the gapped phase of our evolution where $t / t_f$ is small.  In this phase, $B(t) \approx 0$, so we can safely assume that the energy eigenstates remain diagonal in the $\sigma^x$ basis, with ground state $\ket{-}$ and excited state $\ket{+}$, and do not change such that:
\begin{equation}
\langle \pm | \dot{\rho} | \pm \rangle = \frac{d}{dt} \left( \langle \pm | \rho | \pm \rangle \right) \equiv \dot{\rho}_{\pm \pm}  \ .
\end{equation}
The resulting equations for the density matrix elements for small times are then
\bes
 \label{eqt:2qubit}
\begin{eqnarray}
\dot{\rho}_{++}(t) &=& | \langle + | \sigma^z | - \rangle |^2 \left(  \gamma(-\Delta(t)) \rho_{--} -  \gamma(\Delta(t)) \rho_{++} \right) = \gamma_0(\Delta(t))   \left( e^{-\beta \Delta(t)} \rho_{--} - \rho_{++} \right) \ , \\
\rho_{--}(t) &=& 1 - \rho_{++}(t)\ ,
\end{eqnarray}
\ees
where we have used that $\langle + | \sigma^z | + \rangle = \langle - | \sigma^z | - \rangle = 0$, $ \langle + | \sigma^z | - \rangle = 1$, and  $\left[ H_S, \rho_S \right] \approx 0$.  Interpreting Eq.~\eqref{eqt:2qubit} as a rate equation, we see that $\gamma(+\Delta)$ is associated with relaxation from the higher energy state to the lower energy state, and $\gamma (- \Delta (t))$ is associated with the excitation from the lower energy state to the higher energy state. Note that at $t = 0$, we assume that the system is initialized in a thermal state such that $\rho_{++}/ \rho_{--} = e^{- \beta \Delta(0)} \approx 10^{-11}$. Since the gap remains relatively large during the gapped phase, the change in the initial thermal state is minimal.
\subsection{$N$ qubits}
%
For the $N$-qubit case, we can simply generalize our arguments for the single qubit case.  For sufficiently small times the system is diagonal in the $\sigma^x$ basis, and the lowest lying energy states can be considered to be single spin flips in the $x$ direction.  Furthermore, the gap between the ground state and the first excited states is
$\Delta(t) = 2 A(t)$ [Eq.~\eqref{eq:gap}],
which is large in our case.  We can restrict ourselves to only these low lying energy states since higher excited states will have twice the gap to the ground state, and so their contribution will be further suppressed at short times.  

Now, using Eq.~\eqref{eq:flips} and the KMS condition [Eq.~\eqref{eq:KMS-ex}] we find a similar set of equations as in the single qubit case:
\bes
\begin{eqnarray}
\label{eq:J6a}
\dot{\rho}_{ii} &\approx& \sum_{\alpha \beta} \gamma_{\alpha \beta} (\omega_{i 0})\langle 0 | \sigma^z_{\beta} | i \rangle \langle i |  \sigma^z_{\alpha}  | 0 \rangle \left(- \rho_{ii} + e^{- \beta \omega_{i 0}}  \rho_{00} \right) =   \gamma_{i i} (\omega_{i 0}) \left( e^{- \beta \omega_{i 0}}  \rho_{00} - \rho_{ii} \right) \\
\rho_{00} &\approx&1 - N \rho_{ii} \ ,
\end{eqnarray}
\ees
where we have again assumed that the initial state is the thermal state and the gap is large.  

\bibliographystyle{apsrev4-1}

%

\end{document}